\documentclass[aps,11pt,preprintnumbers,nofootinbib,floatfix]{revtex4}

\usepackage{graphicx}
\usepackage{epsfig}
\usepackage{dcolumn}
\usepackage{subfigure}
\usepackage{multirow}

\def\be{\begin{equation}}
\def\ee{\end{equation}}
\def\bea{\begin{eqnarray}}
\def\eea{\end{eqnarray}}

\begin{document}

\title{A flipped $U(1)_R$ extension of the Standard Model}


\author{Cao H. Nam}
\email{caohoangnam@duytan.edu.vn}  
\affiliation{Institute of Research and Development, Duy Tan 
University, Da Nang 550000, Vietnam}
\date{\today}

\begin{abstract}%
In this work, we study an extension of the Standard Model (SM) based on the gauge symmetry $SU(3)_C\times SU(2)_L\times U(1)_{Y'}\times U(1)_R$ where only the right-handed fermions have nonzero $U(1)_R$ charges and the $U(1)_Y$ weak hypercharge of the SM is identified as a combination of the $U(1)_{Y'}$ and $U(1)_R$ charges. The gauge charge assignment of the fields is constrained by the conditions of the anomaly cancellation and the gauge invariance of the Yukawa couplings. The light neutrino masses are generated via the type-I seesaw mechanism where the Majorana masses of the right-handed neutrinos are related to the $U(1)_{Y'}\times U(1)_R$ symmetry breaking scale. Then, we discuss the constraints on the free parameters of the model from the various current experiments, such as precision measurement of the total $Z$ width, $\rho$ parameter, atomic parity violation of Cesium, LEP and LHC bounds. In addition, we investigate the potential of probing for the signal of the new neutral gauge boson based on the forward-backward asymmetry for the process $e^+e^-\rightarrow\mu^+\mu^-$ which is the most sensitive mode at ILC.
\end{abstract}

\maketitle

\section{Introduction}
Tiny but non-vanishing neutrino masses and their mixing \cite{Fukuda1998a,Fukuda1998b,Fukuda1998c} have provided a solid observational evidence which hints at new physics beyond the Standard Model (SM). Adding a new $\mathrm{U}(1)$ symmetry provides a minimal and well-motivated extension of the SM at which it predicts electrically neutral, colorless vector gauge boson. The additional $\mathrm{U}(1)$ symmetry could arise from grand unified models \cite{Langacker1981,Rosner1982a,Rosner1982b,Rosner1984,Rosner1986,Rizzo1989}, from the left-right symmetric models \cite{Mohapatra1975,Mohapatra1980,Mohapatra1981}, from a more fundamental structure of the spacetime \cite{Nam2019a,Nam2019b}, or from various simple extra $\mathrm{U}(1)$ symmetries \cite{Mo-Marshak1980,He-Joshi1991,Rajpoot1993,Deshpande2001,Appelquist2003,Tait2004,Freitas2004,Nath2004a,Nath2004b,Nath2005,
Pokorski2006,Khalil2008,Gopalakrishna2008,Masiero2008,VPleitez2009,Orikasa2009,Khalil2010,HSLee2013,
Sugiyama2014,TCYuan2014,Takahashi2015,Das2016,Nomura2016,Nomura2017,Nomura2018,Chao2018,Raut2018,Zakeri2017}. In the $\mathrm{U}(1)$ extensions of the SM, the three SM singlet right-handed neutrinos arise naturally to achieve the anomaly cancellations. Hence, it can give a natural explanation for the light neutrino masses via the type-I seesaw mechanism at which the Majorana  masses of the right-handed neutrinos are related to the symmetry breaking scale of the new $\mathrm{U}(1)$ symmetry.

Of the extensions of the SM with an additional Abelian gauge symmetry is based on the gauge symmetry $SU(3)_C\times SU(2)_L\times U(1)_Y\times U(1)_R$ where the right-handed fermions are charged under the right-handed gauge symmetry $U(1)_R$, whereas the left-handed fermions have no charges under this symmetry \cite{Nomura2016,Nomura2017,Nomura2018,Chao2018}. The charge assignment depending on the chirality of fermions is quite natural due to their chiral nature. The $U(1)_R$ extension of the SM has attracted a lot of attention and is one of the promising candidates for new physics beyond the SM because of reasons. The three right-handed neutrinos are naturally accommodated as a result of the anomaly cancellations and thus can account for the light neutrino masses. This extension can arise from $SO(10)$ grand unified models \cite{Rizzo1989} or from the left-right symmetric models \cite{Mohapatra2003}. The Higgs
vacuum stablility problem can be solved because the running of the Higgs quartic coupling gets modified due to its coupling to new gauge boson \cite{Chao-Gonderinger2012}. In addition, the $U(1)_R$ extension of the SM  provides richer phenomenology and the candidate for dark matter \cite{Chao2018,SJana2019}.

In this paper, we aim to provide a new understanding of $U(1)_R$ extension of the SM by proposing another scenario based on the following gauge symmetry
\begin{eqnarray}
SU(3)_C\times SU(2)_L\times U(1)_{Y'}\times U(1)_R,\label{GSG}
\end{eqnarray}
and then study the constraints consistent with current experimental data as well as the potential of probing for the signal of the corresponding new gauge boson at the colliders. In this scenario, the charge operator is identified by
\begin{eqnarray}
Q=T_3+Y'+X_R,\label{QO}
\end{eqnarray}
where $Y'$ and $X_R$ are the charges associated with $U(1)_{Y'}$ and $U(1)_R$, respectively. By this way, this model is called flipped $U(1)_R$ extension of the SM. Obviously, the weak hypercharge in the SM is related to the charges $Y'$ and $X_R$ as, $Y=Y'+X_R$. The spontaneous symmetry breaking of the gauge group (\ref{GSG}) is performed through two stages. At the first stage, the symmetry $U(1)_{Y'}\times U(1)_R$ would be broken down to the weak hypercharge symmetry $U(1)_Y$, whereas at the second one the electroweak symmetry $SU(2)_L\times U(1)_Y$ would be broken down to the electromagnetic symmetry $U(1)_{\text{em}}$.

This work is organized as follows. In section \ref{model}, we formulate our model where the particle content
and the gauge charge assignment are introduced. In addition, we present the scalar, neutrino, and neutral gauge boson sectors and determine the couplings of gauge bosons to the fermions. In section \ref{BDW}, we discuss the two-body decays of the new neutral gauge boson at the tree level. In section \ref{constr}, we study the constraints on the free parameters of the model from current experimental data. In section \ref{ILC}, we consider the potential of testing the model by studying the deviation of the forward-backward asymmetry for the process $e^+e^-\rightarrow\mu^+\mu^-$ at ILC. Finally, our conclusion is drawn in section \ref{conclu}.

\section{\label{model}Model}
In this section, we formulate our model based on the gauge symmetry (\ref{GSG}) with the content of the fermions and scalars and the gauge charge assignment given in Table \ref{gauge-charge}. The nonzero vacuum expectation value (VEV) of the scalar singlet $\Phi$ would break the symmetry $U(1)_{Y'}\times U(1)_R$ down to $U(1)_Y$, whereas nonzero VEV of the scalar doublet $H$ would break the electroweak symmetry $SU(2)_L\times U(1)_Y$ down to $U(1)_{\text{em}}$.
\begin{table}[!h]
\begin{center}
\begin{tabular}{|c|c|c|c|c|c|c|c|c|}
\hline
$ $  & $Q_{a}$ & $u_{aR}$ & $d_{aR}$ & $L_{a}$ & $\nu_{aR}$ & $e_{aR}$ & $H$ & $\Phi$ \\
\hline
$SU(3)_C$ & $\textbf{3}$ & $\textbf{3}$ & $\textbf{3}$ & $\textbf{1}$ & $\textbf{1}$ & $\textbf{1}$ & $\textbf{1}$ & $\textbf{1}$ \\
\hline
$SU(2)_L$ & $\textbf{2}$ & $\textbf{1}$ & $\textbf{1}$ & $\textbf{2}$ & $\textbf{1}$ & $\textbf{1}$ & $\textbf{2}$ & $\textbf{1}$ \\
\hline
$U(1)_{Y'}$ & $\ \ \ \ \frac{1}{6}\ \ \ \ $ & $\frac{2}{3}-X_H$ & $-\frac{1}{3}+X_H$ & $\ \ \ \ -\frac{1}{2}\ \ \ \ $ & $-X_H$ & $-1+X_H$ & $\frac{1}{2}-X_H$ & $2X_H$ \\
\hline
$U(1)_R$ & $0$ & $X_H$ & $-X_H$ & $0$ & $X_H$ & $-X_H$ & $X_H$ & $-2X_H$ \\
\hline
\end{tabular}
\caption{The charges of the fermions and scalars under the gauge symmetry (\ref{GSG}) where $X_H$ is a free parameter, with $a=1,2,3$ are generation indices.}\label{gauge-charge}
\end{center}
\end{table}

The total Lagrangian is, up to the gauge fixing and ghost terms, given by
\begin{eqnarray}
\mathcal{L}&=&-\frac{1}{4}G_{a\mu\nu}G^{\mu\nu}_a-\frac{1}{4}W_{i\mu\nu}W^{\mu\nu}_i-\frac{1}{4}B_{\mu\nu}B^{\mu\nu}-\frac{1}{4}X_{\mu\nu}X^{\mu\nu}+\sum_{f}\bar{f}i\gamma^\mu D_\mu f\nonumber\\
&&+\left(D_\mu H\right)^\dagger(D^\mu H)+\left(D_\mu\Phi\right)^\dagger(D^\mu\Phi)-V(H,\Phi)+\mathcal{L}_{\text{Y}},\label{tot-Lag}
\end{eqnarray}
where the field strength tensors are defined as
\begin{eqnarray}
G_{a\mu\nu}&=&\partial_\mu G_{a\nu}-\partial_\nu G_{a\mu}-g_sf_{abc}G_{b\mu}G_{c\nu},\nonumber\\
W_{i\mu\nu}&=&\partial_\mu W_{i\nu}-\partial_\nu W_{i\mu}-g\epsilon_{ijk}W_{j\mu}W_{k\nu},\nonumber\\
B_{\mu\nu}&=&\partial_\mu B_\mu-\partial_\nu B_\mu,\nonumber\\
X_{\mu\nu}&=&\partial_\mu X_\nu-\partial_\nu X_\mu,
\end{eqnarray}
correspond to the gauge groups $SU(3)_C$, $SU(2)_L$, $U(1)_{Y'}$, and $U(1)_R$, respectively, the covariant derivative $D_\mu$ is given by
\begin{eqnarray}
D_\mu=\partial_\mu+ig_s\frac{\lambda_a}{2}G_{a\mu}+ig\frac{\sigma_i}{2}W_{i\mu}+ig_1Y'B_\mu+ig_2X_RX_{\mu},\label{covder}
\end{eqnarray}
with $\{g_s,g,g_1,g_2\}$ to be the gauge couplings corresponding to $\{SU(3)_C, SU(2)_L, U(1)_{Y'},U(1)_R\}$, the scalar potential is given by
\begin{eqnarray}
  V(H,\Phi)=\mu^2_1H^\dagger H+\lambda_1(H^\dagger H)^2+\mu^2_2\Phi^\dagger\Phi+\lambda_2(\Phi^\dagger\Phi)^2+\lambda_3(H^\dagger H)(\Phi^\dagger\Phi),
\end{eqnarray}
and the Yukawa interactions read
\begin{eqnarray}
\mathcal{L}_{\text{Y}}=-h^e_{ab}\bar{L}_aHe_{bR}-h^\nu_{ab}\bar{L}_a\widetilde{H}\nu_{bR}-h^d_{ab}\bar{Q}_aHd_{bR}-h^u_{ab}\bar{Q}_a\widetilde{H}u_{bR}-h^M_{ab}\bar{\nu}^C_{aR}\nu_{bR}\Phi+\textrm{H.c},
\end{eqnarray}
with $\widetilde{H}=i\sigma_2H^*$. For simplicity, the matrix $h^M $ is considered to be diagonal without loss of generality due to the phase redefinitions of the fields. Note that, in general the gauge boson $B$ can mix
kinetically with the one $X$, but we assume that it is negligible in this work.

After the first stage of the spontaneous symmetry breaking, $U(1)_{Y'}\times U(1)_R\rightarrow U(1)_Y$, one can identify a combination of the gauge bosons $B$ and $X$ corresponding to the zero mass as the gauge field associated with the weak hypercharge symmetry $U(1)_Y$. (Whereas, another combination corresponds to the nonzero mass.) Using this combination and the covariant derivative (\ref{covder}), we can find the relation between the gauge coupling $g'$ of the symmetry $U(1)_Y$ in terms of $g_1$ and $g_2$ as
\begin{eqnarray}
\frac{1}{g'^2}=\frac{1}{g^2_1}+\frac{1}{g^2_2}.
\end{eqnarray}
From this relation and Table \ref{gauge-charge}, we can see that in the limit $g_2\rightarrow\infty$ and $X_H\rightarrow0$, the symmetry $U(1)_{Y'}$ should become $U(1)_{Y}$ and the gauge boson of $U(1)_R$ should decouple to the SM particles.

\subsection{Anomaly cancellation}
We can list the nontrivial anomalies associated with the gauge symmetry (\ref{GSG}) as follows: $[SU(3)_C]^2U(1)_{Y'}$, $[SU(3)_C]^2U(1)_R$, $[SU(2)_L]^2U(1)_{Y'}$, $[SU(2)_L]^2U(1)_R$, $[U(1)_{Y'}]^2U(1)_R$, $[U(1)_R]^2U(1)_{Y'}$, $[U(1)_{Y'}]^3$, $[U(1)_R]^3$, $[\text{Gravity}]^2U(1)_{Y'}$, and $[\text{Gravity}]^2U(1)_R$. All of these anomalies must vanish for the consistent theory. Note that, the anomaly $[SU(2)_L]^2U(1)_R$ at which only the left-handed fermions contribute vanishes automatically due to the fact that the left-handed fermions have no charge under $U(1)_R$. The conditions of the anomaly cancellation lead to the following equations
\begin{eqnarray}
2Y'_Q-Y'_{u_R}-Y'_{d_R}&=&0,\nonumber\\
X_{u_R}+X_{d_R}&=&0,\nonumber\\
3Y'_Q+Y'_L&=&0,\nonumber\\
3\left(Y'_{u_R}\right)^2X_{u_R}+3\left(Y'_{d_R}\right)^2X_{d_R}+\left(Y'_{e_R}\right)^2X_{e_R}+\left(Y'_{\nu_R}\right)^2X_{\nu_R}&=&0,\nonumber\\
3\left(X_{u_R}\right)^2Y'_{u_R}+3\left(X_{d_R}\right)^2Y'_{d_R}+\left(X_{e_R}\right)^2Y'_{e_R}+\left(X_{\nu_R}\right)^2Y'_{\nu_R}&=&0,\nonumber\\
6\left(Y'_{Q}\right)^3+2\left(Y'_{L}\right)^3-3\left(Y'_{u_R}\right)^3-3\left(Y'_{d_R}\right)^3-\left(Y'_{e_R}\right)^3-\left(Y'_{\nu_R}\right)^3&=&0,\nonumber\\
3\left(X_{u_R}\right)^3+3\left(X_{d_R}\right)^3+\left(X_{e_R}\right)^3+\left(X_{\nu_R}\right)^3&=&0,\nonumber\\
6Y'_Q+2Y'_L-3Y'_{u_R}-3Y'_{d_R}-Y'_{e_R}-Y'_{\nu_R}&=&0,\nonumber\\
3X_{u_R}+3X_{d_R}+X_{e_R}+X_{\nu_R}&=&0,\label{anc}
\end{eqnarray}
where $Y'_f$ and $X_f$ refer to the $U(1)_{Y'}$ and $U(1)_R$ charges of the fermion $f$, respectively, and we have used the fact $X_Q=X_L=0$. The second and last lines of Eq. (\ref{anc}) imply the following relation
\begin{eqnarray}
X_{u_R}=-X_{d_R},\ \ \ \ X_{\nu_R}=-X_{e_R},\label{rhf-rel}
\end{eqnarray}
by which the seventh line automatically satisfies. Furthermore, the $U(1)_{Y'}$ and $U(1)_R$ charges of the fermions are constrained by the relation (\ref{QO}) which leads to
\begin{eqnarray}
Y'_Q=\frac{1}{6},\ \ \ \ Y'_L&=&-\frac{1}{2},\ \ \ \ Y'_{u_R}=\frac{2}{3}-X_{u_R},\nonumber\\
Y'_{d_R}=-\frac{1}{3}-X_{d_R},\ \ \ \ Y'_{\nu_R}&=&-X_{\nu_R},\ \ \ \ Y'_{e_R}=-1-X_{e_R}.\label{elchr-rels}
\end{eqnarray}
With the relations (\ref{rhf-rel}) and (\ref{elchr-rels}), it is easily to see that the first, third and eighth lines of Eq. (\ref{anc}) automatically satisfy all. Whereas, the fourth, firth and sixth lines all lead to
\begin{eqnarray}
X_{u_R}=X_{\nu_R}.
\end{eqnarray}
In addition to the conditions of the anomaly cancellation and the relation (\ref{elchr-rels}), we have other constraints from the conditions of the gauge invariance of the Yukawa interactions given by
\begin{eqnarray}
-Y'_L+Y'_H+Y'_{e_R}&=&0,\ \ \ \ X_H+X_{e_R}=0,\nonumber\\
-Y'_L-Y'_H+Y'_{\nu_R}&=&0,\ \ \ \ -X_H+X_{\nu_R}=0,\nonumber\\
-Y'_L+Y'_H+Y'_{d_R}&=&0,\ \ \ \ X_H+X_{d_R}=0,\nonumber\\
-Y'_L-Y'_H+Y'_{u_R}&=&0,\ \ \ \ -X_H+X_{u_R}=0,\nonumber\\
2Y'_{\nu_R}+Y'_{\Phi}&=&0,\ \ \ \ 2X_{\nu_R}+X_\Phi=0,
\end{eqnarray}
where $Y'_H(Y'_\Phi)$ and $X_H(X_Q)$ are the $U(1)_{Y'}$ and $U(1)_R$ charges of the scalar field $H$($\Phi$). Then, we can obtain the $U(1)_{Y'}$ and $U(1)_R$ charges for the fermions and scalar fields in terms of a free parameter $X_H$ as showed in Table \ref{gauge-charge}.

\subsection{Scalar sector}

After the spontaneous symmetry breaking, the doublet and singlet scalars are parameterized as
\begin{equation}
H = \left(
\begin{array}{c}
w^+(x) \\
\frac{v+h(x)+iz(x)}{\sqrt{2}} \\
\end{array}
\right),\ \ \ \ \Phi=\frac{v'+h'(x)+iz'(x)}{\sqrt{2}}.
\end{equation}
Here, VEVs $v$ and $v'$ satisfy the conditions of the potential minimization as
\begin{eqnarray}
\frac{\partial V}{\partial v}=\frac{\partial V}{\partial v'}=0
\end{eqnarray}
with
\begin{eqnarray}
V=\frac{1}{4}\left[2\left(\mu^2_1v^2+\mu^2_2v'^2\right)+\lambda_1v^4+\lambda_2v'^4+\lambda_3v^2v'^2\right],
\end{eqnarray}
which leads to
\begin{eqnarray}
v^2=2\frac{2\lambda_2\mu^2_1-\lambda_3\mu^2_2}{\lambda^2_3-4\lambda_1\lambda_2},\ \ \ \ v'^2=2\frac{2\lambda_1\mu^2_2-\lambda_3\mu^2_1}{\lambda^2_3-4\lambda_1\lambda_2}.
\end{eqnarray}
The $CP$-odd fields $w^+(x)$, $z(x)$ and $z'(x)$ would become Nambu-Goldstone bosons which are eaten
by the charged and two neutral gauge bosons. The $CP$-even fields mixes together with the Lagrangian of their mass determined from the mass terms in the scalar potential as
\begin{eqnarray}
 \mathcal{L}_{\text{S-mass}}=\frac{1}{2}\left(
                                      \begin{array}{cc}
                                        h & h' \\
                                      \end{array}
                                    \right)\left(
                                             \begin{array}{cc}
                                               2\lambda_1v^2 & \lambda_3vv' \\
                                               \lambda_3vv' & 2\lambda_2v'^2 \\
                                             \end{array}
                                           \right)\left(
                                                    \begin{array}{c}
                                                      h \\
                                                      h' \\
                                                    \end{array}
                                                  \right).
\end{eqnarray}
The mass eigenstates are obtained as
\begin{eqnarray}
\left(%
\begin{array}{cc}
  h_1 \\
  h_2 \\
\end{array}%
\right)=\left(%
\begin{array}{cc}
  c_\alpha & -s_\alpha \\
  s_\alpha & c_\alpha \\
\end{array}%
\right)\left(%
\begin{array}{cc}
  h \\
  h' \\
\end{array}%
\right),
\end{eqnarray} 
corresponding to the following masses
\begin{equation}
  m^2_{h_1,h_2}=\lambda_1v^2+\lambda_2v'^2\mp\left[\left(\lambda_1v^2-\lambda_2v'^2\right)^2+\lambda^2_3v^2v'^2\right]^{1/2}.
\end{equation}
Here, the light Higgs boson $h_1$ is identified as the SM Higgs. The mixing angle $\alpha$ between the physical states $h_1$ and $h_2$ is determined by
\begin{eqnarray}
\sin2\alpha=\frac{2\lambda_3vv'}{m^2_{h_2}-m^2_{h_1}},
\end{eqnarray}
which suggests that for $\lambda_3=0$ there is no the mixing between them. Due to this mixing, the SM couplings of $h_1$ to the SM fermions and gauge bosons get modified, which is constrained by the measurements of the Higgs production cross section and its decay branching ratio at LHC leading to $s_\alpha\lesssim0.2$ \cite{Tanabashi2018} from which we obtain the following bound
\begin{eqnarray}
\left|\frac{\lambda_3vv'}{\lambda_1v^2-\lambda_2v'^2}\right|\lesssim0.426.\label{lam3-constr}
\end{eqnarray}

\subsection{Neutrino masses}
The fermions acquire the Dirac masses through VEV of the scalar doublet $H$ at the electroweak symmetry breaking scale $v$. In addition, the right-handed neutrinos get the Majorana masses through VEV of the scalar singlet $\Phi$ at the $U(1)_{Y'}\times U(1)_R$ symmetry breaking scale $v'$ which is (much) larger than $v$. The charge sector is the same the SM one, since here we are only interested in the neutral sector. From the neutrino Yukawa interaction terms, we obtain the Lagrangian of neutrino masses as
\begin{eqnarray}
\mathcal{L}_{\text{N-mass}} &=& -\frac{1}{2}\left(%
\begin{array}{cc}
  \bar{\nu}_L & \bar{\nu}^C_R \\
\end{array}%
\right)
\left(%
\begin{array}{cc}
  0 & \frac{h^\nu v}{\sqrt{2}} \\
  \frac{(h^\nu)^Tv}{\sqrt{2}} & \sqrt{2}h^Mv' \\
\end{array}%
\right)
\left(%
\begin{array}{c}
  \nu^C_L \\
  \nu_R \\
\end{array}
\right)+\textrm{H.c.},
\end{eqnarray}
By diagonalizing the mass matrix of the neutrinos, it leads to the following mass eigenvalues
\begin{eqnarray}
m_{\nu'_L}&\approx&-U_{_{MNS}}h^\nu\left(h^M\right)^{-1}\left(h^\nu\right)^T\frac{v^2}{2\sqrt{2}v'}U^\dagger_{_{MNS}},\nonumber\\
m_{\nu'_R}&\approx&\sqrt{2}h^Mv',
\end{eqnarray}
corresponding to the following eigenstates
\begin{eqnarray}
\left(\begin{array}{c}
\nu'^C_L \\
\nu'_R \\
\end{array}
\right) &\approx&\left(
\begin{array}{cc}
U_{_{MNS}} & -\delta^\dagger \\
\delta & 1\\
\end{array}\right)\left(\begin{array}{c}
\nu^C_L \\
\nu_R \\
\end{array}
\right),
\end{eqnarray}
where $U_{_{MNS}}$ is the Maki-Nakagawa-Sakata (MNS) matrix determined by the current neutrino oscillation data \cite{Tanabashi2018} and $\delta=\left(h^M\right)^{-1}\left(h^\nu\right)^Tv/2v'$. The eigenstates $\nu'_L$ are identified as the observed light neutrinos, whereas the eigenstates $\nu'_R$ are the heavy neutrinos. Because the sub-eV neutrino mass scale leads to that the mixing parameter $\delta$ is extremely small, we have the approximation, $\nu'_L\approx U_{_{MNS}}\nu_L$ and $\nu'_R\approx\nu_R$.

\subsection{Neutral gauge boson sector}
The spontaneous symmetry breaking $SU(2)_L\times U(1)_{Y'}\times U(1)_R$ down to $U(1)_{\text{em}}$ gives the masses to the gauge bosons. The SM charge sector of the gauge bosons does not get modified in this work, hence we focus the neutral sector only. The mass matrix of the neutral gauge bosons in the basis $(W_{3\mu}, B_\mu, X_\mu)$ is given by
\begin{eqnarray}
M^2=\left(%
\begin{array}{ccc}
  \frac{g^2v^2}{4} & -\frac{gg_1(1-2X_H)v^2}{4} & -\frac{gg_2X_Hv^2}{2} \\
  -\frac{gg_1(1-2X_H)v^2}{4} & \frac{g^2_1}{4}\left[(1-2X_H)^2v^2+16X^2_H{v'}^2\right] & \frac{g_1g_2X_H}{2}\left[(1-2X_H)v^2-8X_Hv'^2\right] \\
  -\frac{gg_2X_Hv^2}{2} & \frac{g_1g_2X_H}{2}\left[(1-2X_H)v^2-8X_Hv'^2\right] & g^2_2X^2_H(v^2+4v'^2) \\
\end{array}%
\right)
\end{eqnarray} 
By diagonalizing this mass matrix, one can find the mass eigenvalues as, $VM^2_{\text{NGB}}V^T=\text{Diag}(M^2_Z,M^2_{Z'},0)$, where the matrix $V$ is given by
\begin{eqnarray}
V=\left(%
\begin{array}{ccc}
 c_\beta  & -s_\beta &  0\\
 s_\beta & c_\beta & 0 \\
 0 & 0 & 1 \\ 
\end{array}%
\right)\left(%
\begin{array}{ccc}
 \frac{g\sqrt{g^2_1+g^2_2}}{\sqrt{g^2_1g^2_2+g^2(g^2_1+g^2_2)}}  & 0 &  -\frac{g_1g_2}{\sqrt{g^2_1g^2_2+g^2(g^2_1+g^2_2)}}\\
 0 & 1 & 0 \\
 \frac{g_1g_2}{\sqrt{g^2_1g^2_2+g^2(g^2_1+g^2_2)}} & 0 &  \frac{g\sqrt{g^2_1+g^2_2}}{\sqrt{g^2_1g^2_2+g^2(g^2_1+g^2_2)}} \\ 
\end{array}%
\right)\left(%
\begin{array}{ccc}
 1  & 0 &  0\\
 0 & \frac{g_1}{\sqrt{g^2_1+g^2_2}} & -\frac{g_2}{\sqrt{g^2_1+g^2_2}} \\
 0 & \frac{g_2}{\sqrt{g^2_1+g^2_2}} & \frac{g_1}{\sqrt{g^2_1+g^2_2}} \\ 
\end{array}%
\right).
\end{eqnarray}
The mass basis is related to the basis $(W_{3\mu}, B_\mu, X_\mu)$ as, $(Z_{\mu},Z'_{\mu},A_\mu)^T=V(W_{3\mu}, B_\mu, X_\mu)^T$. The masses of the physical gauge bosons $Z_{\mu}$ and $Z'_{\mu}$ is given by
\begin{eqnarray}
M^2_{Z}&=&\frac{1}{4}(g^2+g'^2)v^2-\frac{g^2+g'^2}{64X^2_H}\left[\frac{2(1+t^2)X_H-1}{1+t^2}\right]^2\frac{v^4}{v'^2}+\mathcal{O}(v^4/v'^4),\\
M^2_{Z'}&=&4(g^2_1+g^2_2)X^2_Hv'^2+\frac{(g^2_1+g^2_2)}{4}\left[\frac{2(1+t^2)X_H-1}{1+t^2}\right]^2v^2+\mathcal{O}(v^2/v'^2),
\end{eqnarray}
and the mixing angle $\beta$ between them is determined by
\begin{eqnarray}
\tan2\beta&=&\frac{t\left[2(1+t^2)X_H-1\right]}{8s_WX^2_H(1+t^2)^2}\frac{v^2}{v'^2}+\mathcal{O}(v^4/v'^4),\nonumber\\
&=&2s_W\frac{\left[2(1+t^2)X_H-1\right]}{t}\frac{M^2_{Z}}{M^2_{Z'}}+\mathcal{O}(M^4_{Z}/M^4_{Z'}),
\end{eqnarray}
where $t\equiv g_2/g_1$ and  $s_W\equiv\sin\theta_W$ with $\theta_W$ to be Weinberg angle. The physical state $Z$ is identified as the SM neutral massive boson and another physical state $Z'$ is the heavy new gauge boson predicted by the model.

\subsection{Fermion-gauge boson couplings}
In order to derive the couplings between the gauge bosons and the fermions, let first us expand the Lagrangian for the fermions in (\ref{tot-Lag}) as
\begin{eqnarray}
\sum_{f}\bar{f}i\gamma^\mu D_\mu f&=&\sum_{f_L}\bar{f}_Li\gamma^\mu\left(\partial_\mu+ig\frac{\sigma_i}{2}W_{i\mu}+ig_1Y'_{f_L}B_\mu+ig_2X_{f_L}X_\mu\right)f_L\nonumber\\
&&+\sum_{f_R}\bar{f}_Ri\gamma^\mu\left(\partial_\mu+ig_1Y'_{f_R}B_\mu+ig_2X_{f_R}X_\mu\right)f_R,
\end{eqnarray}
where we have dropped the terms relating to the symmetry $SU(3)_C$. From this and the relation $(W_{3\mu}, B_\mu, X_\mu)^T=V^T(Z_{\mu},Z'_{\mu},A_\mu)^T$, once can find the couplings of the physical neutral gauge bosons to the fermions as
\begin{eqnarray}
\mathcal{L}_{NC}&=&-\left[C^{Z}_{\nu'_L}\bar{\nu}'_L\gamma^\mu\nu'_L+C^{Z}_{\nu'_R}\bar{\nu}'_R\gamma^\mu\nu'_R\right]Z_\mu-\left[C^{Z'}_{\nu'_L}\bar{\nu}'_L\gamma^\mu\nu'_L+C^{Z'}_{\nu'_R}\bar{\nu}'_R\gamma^\mu\nu'_R\right]Z'_\mu\nonumber\\
&&-\sum_f\left[\bar{f}\gamma^\mu(C^{Z}_{V,f}+C^{Z}_{A,f}\gamma_5)fZ_{\mu}+\bar{f}\gamma^\mu(C^{Z'}_{V,f}+C^{Z'}_{A,f}\gamma_5)fZ'_{\mu}\right],
\end{eqnarray}
where $f$ only refers to the charged fermions and
\begin{eqnarray}
C^{Z}_{\nu'_L}&=&\frac{g}{2c_W}\left(c_\beta+\frac{s_Ws_\beta}{t}\right),\nonumber\\
C^{Z'}_{\nu'_L}&=&\frac{g}{2c_W}\left(s_\beta-\frac{s_Wc_\beta}{t}\right),\nonumber\\
C^{Z}_{\nu'_R}&=&gt_WX_Hs_\beta\frac{1+t^2}{t},\nonumber\\
C^{Z'}_{\nu'_R}&=&-gt_WX_Hc_\beta\frac{1+t^2}{t},\nonumber\\
C^{Z}_{V,f}&=&\frac{gc_\beta}{2c_W}\left[T_{3}(f_L)-2Q_fs^2_W\right]-gt_Ws_\beta\frac{(Y'_{f_L}+Y'_{f_R})-t^2(X_{f_L}+X_{f_R})}{2t},\nonumber\\
C^{Z}_{A,f}&=&-\frac{gc_\beta}{2c_W}T_{3}(f_L)-gt_Ws_\beta\frac{(Y'_{f_R}-Y'_{f_L})-t^2(X_{f_R}-X_{f_L})}{2t},\nonumber\\
C^{Z'}_{V,f}&=&\frac{gs_\beta}{2c_W}\left[T_{3}(f_L)-2Q_fs^2_W\right]+gt_Wc_\beta\frac{(Y'_{f_L}+Y'_{f_R})-t^2(X_{f_L}+X_{f_R})}{2t},\nonumber\\
C^{Z'}_{A,f}&=&-\frac{gs_\beta}{2c_W}T_{3}(f_L)+gt_Wc_\beta\frac{(Y'_{f_R}-Y'_{f_L})-t^2(X_{f_R}-X_{f_L})}{2t},
\end{eqnarray}
with $c_W\equiv\cos\theta_W$ and $t_W\equiv\tan\theta_W$. The mixing does not change the electromagnetic couplings and the charged currents for the quarks. The charged currents for the leptons is given by
\begin{eqnarray}
\mathcal{L}_{CC}&=&-\frac{g}{\sqrt{2}}\bar{l}_L\gamma^\mu U^\dagger_{MNS}\nu'_LW^-_\mu+\textrm{H.c.}.
\end{eqnarray}
 
\section{\label{BDW}Partial decay widths and branching ratios}
We study the two-body decays of the extra gauge boson $Z'$ predicted by our model. In this work, we are interested in the case $M_{Z'}<m_{\nu'_R}$ and $M_{Z'}<m_{h_2}$ and consider the tree-level processes. Due to the gauge boson $Z'$ having the couplings with leptons and quarks, $Z'$ can decay to the fermion pairs $Z'\rightarrow\bar{f}f$ with the decay width given by
\begin{eqnarray}
\Gamma(Z'\rightarrow\bar{f}f)=\frac{N_C(f)M_{Z'}}{12\pi}\sqrt{1-\frac{4m^2_f}{M^2_{Z'}}}\left
\{\left[\left(C^{Z'}_{V,f}\right)^2+\left(C^{Z'}_{A,f}\right)^2\right]\left(1+\frac{2m^2_f}{M^2_{Z'}}\right)-6\left(C^{Z'}_{A,f}\right)^2\frac{m^2_f}{M^2_{Z'}}\right\}.
\end{eqnarray}
In addition, due to the mixing between the SM gauge boson $Z$ and the new gauge boson $Z'$, there are other modes which $Z'$ can decay to the pairs $W^+W^-$ and $Zh_1$, with decay widths given by \cite{Enberg2016}
\begin{eqnarray}
\Gamma(Z'\rightarrow W^+W^-)&=&(gc_Ws_\beta)^2\frac{M^5_{Z'}}{192\pi M^4_W}\left(1-\frac{4M^2_W}{M^2_{Z'}}\right)^{3/2}\left(1+\frac{20M^2_W}{M^2_{Z'}}+\frac{12M^4_W}{M^4_{Z'}}\right),\nonumber\\
\Gamma(Z'\rightarrow Zh_1)&=&\frac{\left(g_{_{Z'Zh_1}}\right)^2M_{Z'}}{192\pi M^2_{Z}}\left[1-\frac{2m^2_{h_1}-10M^2_{Z}}{M^2_{Z'}}+\frac{(m^2_{h_1}-M^2_{Z})^2}{M^4_{Z'}}\right]\nonumber\\
&&\times\left[1-\frac{2(m^2_{h_1}+M^2_{Z})}{M^2_{Z'}}+\frac{(m^2_{h_1}-M^2_{Z})^2}{M^4_{Z'}}\right]^{1/2},
\end{eqnarray}
where the mass dimension coupling $g_{_{Z'Zh_1}}$ is given by
\begin{eqnarray}
g_{_{Z'Zh_1}}&=&gM_W\left[\frac{c^2_Wc_\alpha}{2}s_{2\beta}-t^2_W\left[(1-2X_H)^2c_\alpha-16X^2_H\frac{v'}{v}s_\alpha\right]\left(\frac{s_W}{t}c_{2\beta}+\frac{1-t^2s^2_W}{2t^2}s_{2\beta}\right)\right.\nonumber\\
&&\left.+4X_H^2t^2_W\left(c_\alpha-4\frac{v'}{v}s_\alpha\right)\left(ts_Wc_{2\beta}-\frac{t^2-s^2_W}{2}s_{2\beta}\right)+2X_Ht_Wc_\alpha\left(tc_Wc_{2\beta}+\frac{s_{2\theta_W}}{2}s_{2\beta}\right)\right.\nonumber\\
&&\left.+2t^2_WX_H\left((1-2X_H)c_\alpha+8X_H\frac{v'}{v}s_\alpha\right)\left(\frac{(t^2-1)s_W}{t}c_{2\beta}+(1+s^2_W)s_{2\beta}\right)\right.\nonumber\\
&&\left.-(1-2X_H)t_Wc_\alpha\left(\frac{c_W}{t}c_{2\beta}-\frac{s_{2\theta_W}}{2}s_{2\beta}\right)\right],
\end{eqnarray}
with $s_{2\beta}\equiv\sin2\beta$, $c_{2\beta}\equiv\cos2\beta$, and $s_{2\theta_W}\equiv\sin2\theta_W$.

\begin{figure}[t]
 \centering
\begin{tabular}{cc}
\includegraphics[width=0.45 \textwidth]{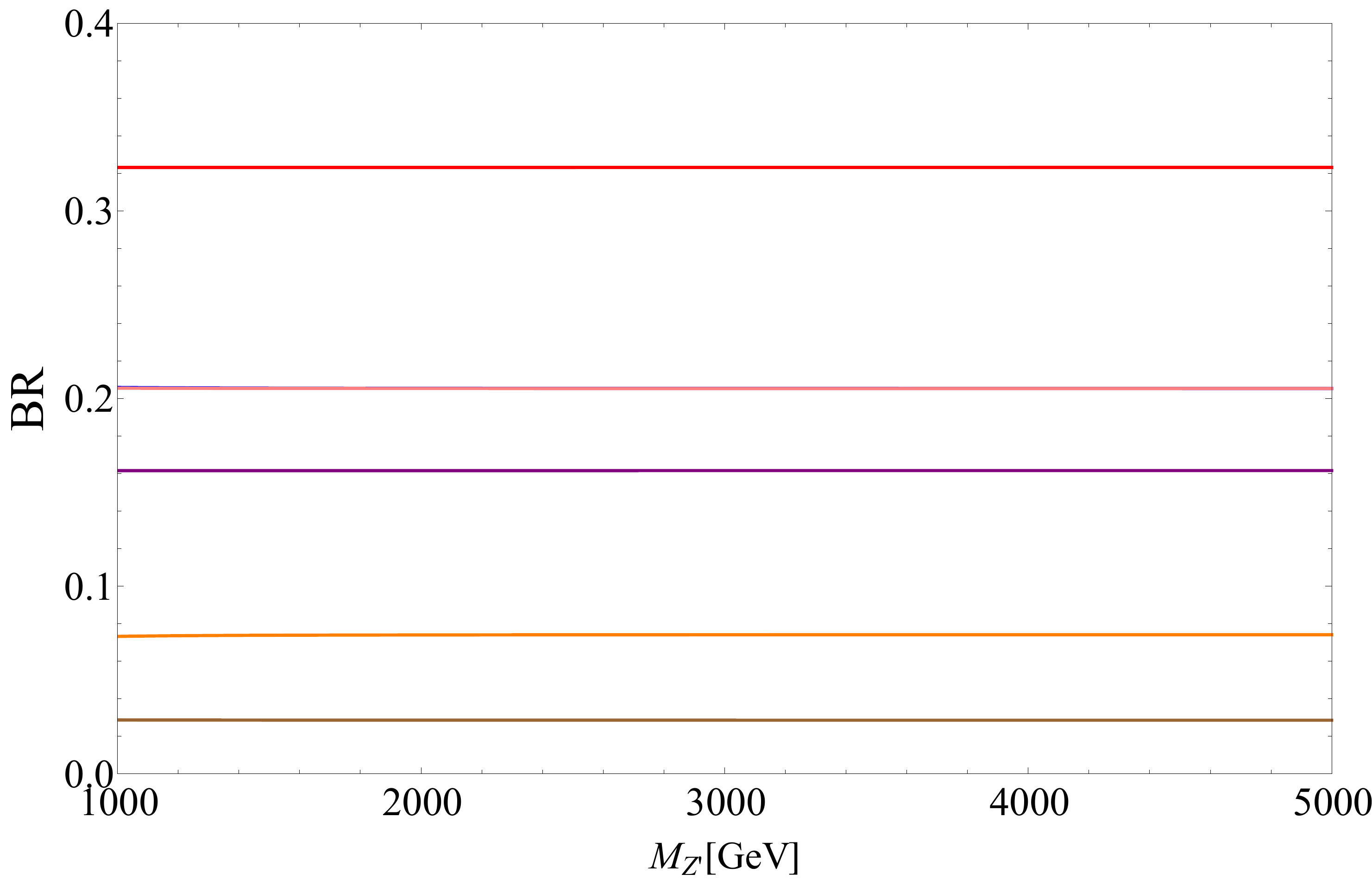}
\hspace*{0.05\textwidth}
\includegraphics[width=0.45 \textwidth]{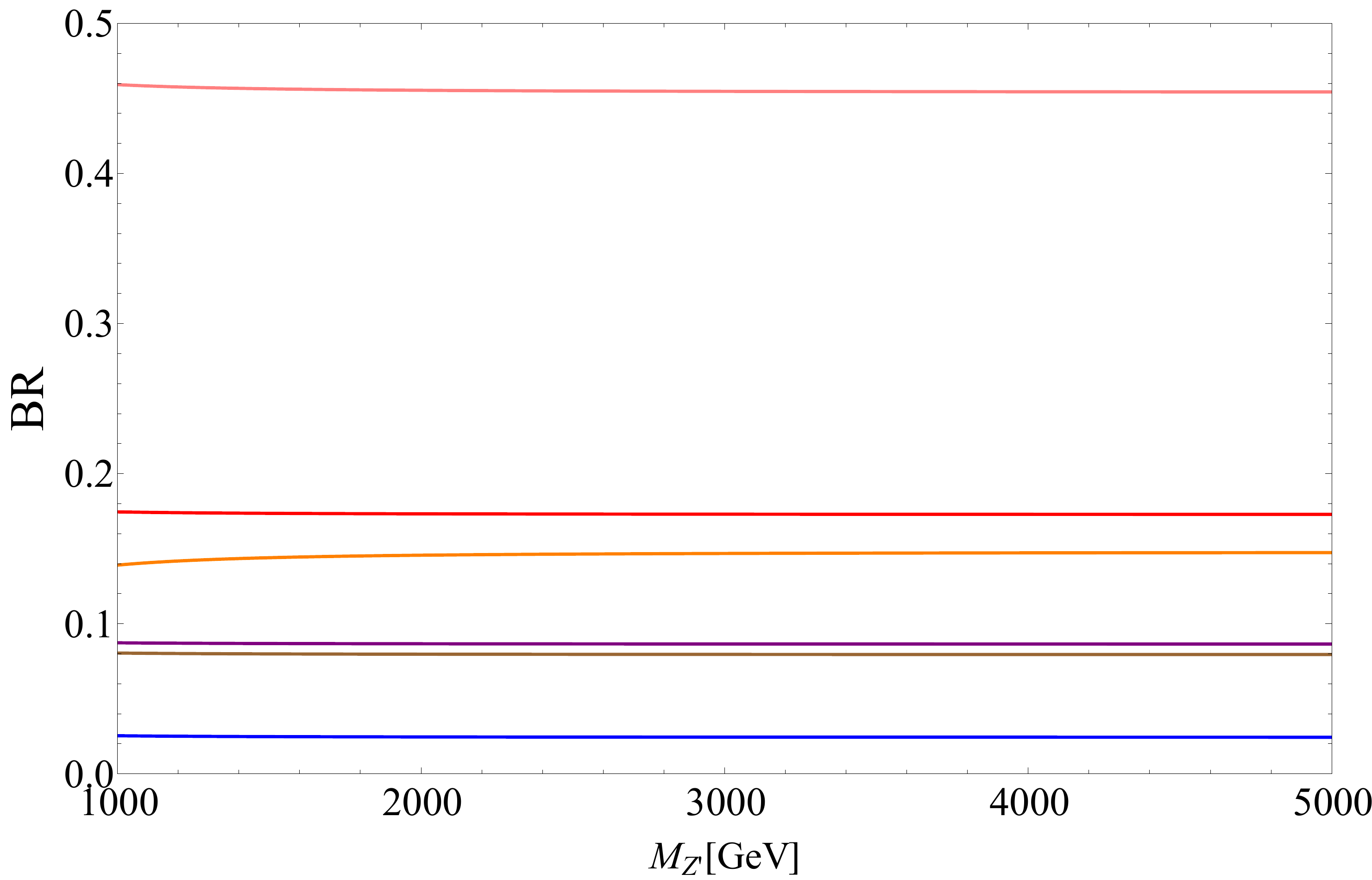}
\end{tabular}
  \caption{The decay branching ratios of the new gauge boson $Z'$ into the two-body final states at $s_\alpha=0.1$. The blue, red, purple, pink, brown, and orange lines correspond to the decay chanels $Z'\rightarrow3\bar{\nu}'_L\nu'_L$, $\bar{l}l$ $(l=e,\mu)$, $\bar{\tau}\tau$, $\bar{q}q$ $(q=u,d,s,c)$, $\bar{b}b$, and $\bar{t}t$. Lef panel: $t=1/2$ and $X_H=1/3$. Right panel: $t=1/3$ and $X_H=-1/2$. }\label{BRs}
\end{figure}

In figure \ref{BRs}, we show the two-body decay branching ratios of the new gauge boson $Z'$ as a function of its mass $M_{Z'}$. We find the values of the branching ratios of $Z'$ decays in the left panel (right panel) of this figure as
\begin{eqnarray}
\text{BR}(Z'\rightarrow3\bar{\nu}'_L\nu'_L)\sim20.5\% (2.5\%),\ && \ \sum_{l=e,\mu}\text{BR}(Z'\rightarrow\bar{l}l)\sim32.3\% (17.3\%),\nonumber\\
\text{BR}(Z'\rightarrow\bar{\tau}\tau)\sim16.2\% (8.7\%),\ && \ \sum_{q=u,d,s,c}\text{BR}(Z'\rightarrow\bar{q}q)\sim20.5\% (45.5\%),\nonumber\\
\text{BR}(Z'\rightarrow\bar{b}b)\sim2.9\% (8.0\%),\  && \ \text{BR}(Z'\rightarrow\bar{t}t)\sim7.4\% (14.6\%),\nonumber\\
\text{BR}(Z'\rightarrow W^+W^-)&\approx&\text{BR}(Z'\rightarrow Zh_1)\sim0.1\% (1.7\%).
\end{eqnarray}
According to this figure, the dominant decay channels depend on the values of the free parameters $t$ and $X_H$. For the left panel of figure \ref{BRs}, the final state with the highest branching ratio is the dilepton modes. Also, the branching ratio of $\sum_{l=e,\mu}Z'\rightarrow\bar{l}l$ is relatively higher than the branching ratio of $\sum_{q=u,d,s,c}Z'\rightarrow\bar{q}q$ in the entire range of the $Z'$ boson mass in consideration, which suggests that search for $Z'$ at LHC can be accessible through a clean dilepton signal. This is due to that the values of the free parameters $t$ and $X_H$ in this case lead to the $Z'$ couplings to leptons large compared to the $Z'$ couplings to quarks. In addition, the ratio $\Gamma_{Z'}/M_{Z'}$ obtained in this case is about $3\%$ which implies a narrow resonance in the invariant mass distribution of the final-state system. On the contrary, for the right panel of figure \ref{BRs}, the decay of $Z'$ is sensitive to the dijet modes compared to dilepton modes. The ratio $\Gamma_{Z'}/M_{Z'}$ obtained in this case is about $47\%$ which means that the $Z'$ resonance should be very broad. Note that, the branching ratios of $Z'\rightarrow W^+W^-$ and $Z'\rightarrow Zh_1$ are quite small and are approximately equal together due to the consequence of Goldstone boson equivalence in the high energy limit.

\section{\label{constr}Constraints on the new physics}
In this section, we are interested in the constraints on the free parameters of the model using various current experiments from which the allowed parameter space is obtained.

\subsection{Precision measurement of the total $Z$
width}

The tree-level mass mixing between the SM gauge boson $Z$ and the new gauge boson $Z'$ would modify both the mass and couplings of $Z$. As a result, it would lead to the corrections to the relevant SM predictions, which should be constrained by the precision electroweak measurements such as the precision measurement of the total $Z$ width. 

First, let us rewrite the $Z$ couplings to the SM fermions in our model as
\begin{eqnarray}
\mathcal{L}_{Z}&=&-\bar{f}\gamma^\mu\left[C^{\text{SM}}_{V,f}\left(1+\delta_{V,f}\right)+C^{\text{SM}}_{A,f}\left(1+\delta_{A,f}\right)\gamma_5\right]fZ_{\mu}-\frac{g}{2c_W}\left(1+\delta_{\nu'_L}\right)\bar{\nu}'_L\gamma^\mu\nu'_LZ_\mu,
\end{eqnarray}
where $f$ only refers to the charged fermions, $C^{\text{SM}}_{V,f}=\frac{g}{2c_W}\left[T_{3}(f_L)-2Q_fs^2_W\right]$ and $C^{\text{SM}}_{A,f}=-gT_3(f_L)/2c_W$ are the SM values for the vector and axial couplings of the SM gauge boson $Z$ to the charged fermions, respectively, the shifts $\delta_{V,f}$ and $\delta_{A,f}$ in these couplings are given by
\begin{eqnarray}
\delta_{V,f}&=&-s^2_W\frac{\left[2(1+t^2)X_H-1\right]\left[(Y'_{f_L}+Y'_{f_R})-t^2(X_{f_L}+X_{f_R})\right]}{\left[T_{3}(f_L)-2Q_fs^2_W\right]t^2}\frac{M^2_{Z}}{M^2_{Z'}}+\mathcal{O}(M^4_{Z}/M^4_{Z'}),\nonumber\\
\delta_{A,f}&=&s^2_W\frac{\left[2(1+t^2)X_H-1\right]\left[(Y'_{f_R}-Y'_{f_L})-t^2(X_{f_R}-X_{f_L})\right]}{t^2T_3(f_L)}\frac{M^2_{Z}}{M^2_{Z'}}+\mathcal{O}(M^4_{Z}/M^4_{Z'}),
\end{eqnarray}
and the shift in the $Z$ coupling to the light neutrinos is given by
\begin{eqnarray}
\delta_{\nu'_L}=s^2_W\frac{\left[2(1+t^2)X_H-1\right]}{t^2}\frac{M^2_{Z}}{M^2_{Z'}}+\mathcal{O}(M^4_{Z}/M^4_{Z'}).
\end{eqnarray}
Then, we can write the total width of the SM gauge boson $Z$ in our model as
\begin{eqnarray}
\Gamma_{Z}=\Gamma^{\text{SM}}_{Z}+\delta\Gamma_{{Z}},
\end{eqnarray} 
where $\Gamma^{\text{SM}}_{Z}$ is the SM prediction and the shift $\delta\Gamma_{{Z}}$ in the total $Z$ width is given by
\begin{eqnarray}
\delta\Gamma_{{Z}}&\simeq&\frac{M_Z}{6\pi}\sum_fN_C(f)\left[\left(C^{\text{SM}}_{V,f}\right)^2\delta_{V,f}+\left(C^{\text{SM}}_{A,f}\right)^2\delta_{A,f}\right]+\frac{M_Z\delta_{\nu'_L}}{4\pi}\left(\frac{g}{2c_W}\right)^2\nonumber\\
&&+\frac{\delta M_Z}{12\pi}\sum_fN_C(f)\left[\left(C^{\text{SM}}_{V,f}\right)^2+\left(C^{\text{SM}}_{A,f}\right)^2\right]+\frac{\delta M_Z}{8\pi}\left(\frac{g}{2c_W}\right)^2,
\end{eqnarray}
with the sum taken over the relevant charged fermions and the shift $\delta M_Z$ in the mass of the SM gauge boson $Z$ given by
\begin{eqnarray}
\delta M_Z=-\frac{s^2_W}{2}\left[\frac{2(1+t^2)X_H-1}{t}\right]^2\frac{M^3_Z}{M^2_{Z'}}+\mathcal{O}(M^4_{Z}/M^4_{Z'}).
\end{eqnarray}
With the experimental value of the total $Z$ width $\Gamma^{\text{exp}}_Z=2.4952\pm0.0023$ GeV and the SM prediction $\Gamma^{\text{SM}}_Z=2.4942\pm0.0008$ GeV \cite{Tanabashi2018}, we impose a constraint $|\delta\Gamma_Z|<0.0025$ GeV from which one can obtain a lower bound on the mass of the new gauge boson $Z'$ as
\begin{eqnarray}
M_{Z'}\gtrsim2.92\times\frac{\Big|0.37+X_H\left[X_H-1.25(1+t^2)+t^2(2+t^2)X_H\right]\Big|^{1/2}}{t} \ \ \text{TeV},
\end{eqnarray}
which depends on the free parameters $t$ and $X_H$.
\subsection{Perturbativity condition}
If the $Z'$ couplings are not too big, the perturbation approximation in the relevant processes is reliable. We impose the following simple perturbativity condition on the ratio of the total $Z'$
width to the $Z'$ gauge boson mass \cite{Zhang2019}
\begin{eqnarray}
\frac{\Gamma_{Z'}}{M_{Z'}}<1,
\end{eqnarray}
where
\begin{eqnarray}
\Gamma_{Z'}=\sum_f\Gamma(Z'\rightarrow\bar{f}f)+\Gamma(Z'\rightarrow W^+W^-)+\Gamma(Z'\rightarrow Zh_1),
\end{eqnarray}
with the sum taken over all fermions except the heavy neutrinos $\nu'_R$ and the partial widths given in the previous section. This condition leads to a constraint for the free parameters $t$ and $X_H$ as
\begin{eqnarray}
0.47-26.85t^2-1.14(1+t^2)X_H+\left[1+t^2(t^2+2)\right]X^2_H\lesssim0.\label{perconst}
\end{eqnarray}

\subsection{$\rho$ parameter}

One of the most important observables, which is used to constrain models of new physics, is $\rho$ parameter defined by
\begin{eqnarray}
\rho=\frac{M^2_W}{c^2_WM^2_{Z}},
\end{eqnarray}
which is equal to one in the SM. At the tree level, the contribution of new physics to $\rho$ parameter comes from the mixing of the SM gauge boson $Z$ to new one $Z'$, which is given by
\begin{eqnarray}
\Delta\rho=\rho-1\simeq s^2_W\left[\frac{2(1+t^2)X_H-1}{t}\right]^2\frac{M^2_Z}{M^2_{Z'}}.
\end{eqnarray}
The experimental value of $\rho$ parameter is given by $\rho=1.00039\pm0.00019$ which is $2\sigma$ above the SM expectation $\rho_{\text{SM}}=1$ \cite{Tanabashi2018}. Since if new physics exits it must satisfy $\Delta\rho<0.00058$ which leads to the following lower bound
\begin{eqnarray}
M_{Z'}\gtrsim3.64\times\frac{|(1+t^2)X_H-0.5|}{t} \ \ \text{TeV}.
\end{eqnarray}

\subsection{Atomic parity violation of Cesium}

In the SM, the gauge boson $Z$ causes the atomic parity violation (APV) characterized in terms of the weak nuclear charge $Q_W$ of a nucleus. In our model, the new gauge boson $Z'$ would lead to an additional contribution to the weak nuclear charge. Currently, the weak nuclear charge of Cesium
has been measured to a precision given by \cite{Porsev09-10,Wood1997,Bennet}
\begin{eqnarray}
Q^{\text{exp}}_W(^{133}_{\phantom{k} 55}\text{Cs})&=&-73.16(29)_{\text{exp}}(20)_{\text{th}},
\end{eqnarray}
which is in agreement with the SM prediction (including electroweak radiative corrections) \cite{Marciano84-85,Marciano90-92} as
\begin{eqnarray}
Q^{\text{th}}_W(^{133}_{\phantom{k} 55}\text{Cs})&=&-73.16(3).
\end{eqnarray}
Thus, this imposes a constraint on the contribution of new physics for the nuclear weak charge of Cesium as, $\left|\Delta Q_W(^{133}_{\phantom{k} 55}\text{Cs})\right|\lesssim0.52$. Using the general calculation for the correction of the weak nuclear charge of given isotope (mediated by a massive gauge boson) in Ref. \cite{Diener2012}, we find $\Delta Q_W(^{133}_{\phantom{k} 55}\text{Cs})$ due to the contribution of the new gauge boson $Z'$ as
\begin{eqnarray}
  \Delta Q_W(^{133}_{\phantom{k} 55}\text{Cs})=-16\left(\frac{M_{Z}}{M_{Z'}}\right)^2\left(\frac{c_W}{g}\right)^2C^{Z'}_{A,e}\left[(2Z+N)C^{Z'}_{V,u}+(Z+2N)C^{Z'}_{V,d}\right],
\end{eqnarray}
with $Z=55$ and $N=78$. Then, one can find a lower bound on the mass of the new gauge boson $Z'$ as
\begin{eqnarray}
M_{Z'}\gtrsim1.34\times\frac{|\left[1+0.2(1+t^2)X_H\right]\left[(1+t^2)X_H-0.5\right]|^{1/2}}{t}\ \ \text{TeV}.
\end{eqnarray}

\subsection{LEP constraint}

The presence of the new gauge boson $Z'$ should lead to the deviations from the SM prediction in the process $e^+e^-\rightarrow l^+l^-$ with $(l=e,\mu,\tau)$, which is constrained by the LEP data \cite{}. By integrating out this heavy gauge boson, one can find a contact interaction for the process $e^+e^-\rightarrow l^+l^-$ which is parametrised by the following effective Lagrangian
\begin{eqnarray}
\mathcal{L}_{\text{eff}}&=&\frac{1}{1+\delta_{el}}\frac{1}{M^2_{Z'}}\left[C^{Z'}_{L,e}\bar{e}_L\gamma_\mu e_L+C^{Z'}_{R,e}\bar{e}_R\gamma_\mu e_R\right]\left[C^{Z'}_{L,l}\bar{l}_L\gamma^\mu l_L+C^{Z'}_{L,l}\bar{l}_R\gamma^\mu l_R\right],\nonumber\\
&=&\frac{1}{1+\delta_{el}}\frac{1}{M^2_{Z'}}\sum_{i,j=L,R}\eta_{ij}\bar{e}_i\gamma_\mu e_i\bar{l}_j\gamma^\mu l_j,
\end{eqnarray}
where $\delta_{el}=1(0)$ for $l=e$ ($l\neq e$), the $Z'$ couplings to the left- and right-handed leptons are given by
\begin{eqnarray}
C^{Z'}_{L,l}&=&\frac{g}{2c_{W}}\left[s_\beta(2s^2_W-1)-\frac{s_Wc_\beta}{t}\right],\nonumber\\
C^{Z'}_{R,l}&=&gt_W\left[s_Ws_\beta+c_\beta\frac{(1+t^2)X_H-1}{t}\right],
\end{eqnarray}
and $\eta_{ij}=C^{Z'}_{i,e}C^{Z'}_{j,l}$. By fitting this contact interaction to the relevant LEP data \cite{Schael2013}, it leads to
\begin{eqnarray}
\frac{2\sqrt{2\pi}M_{Z'}}{\sqrt{\left(C^{Z'}_{L,e}\right)^2+\left(C^{Z'}_{R,e}\right)^2}}\gtrsim24.6\ \ \text{TeV},
\end{eqnarray}
for $(1+t^2)X_H<1$ which corresponds to the case $\eta_{LR},\eta_{RL}>0$, and
\begin{eqnarray}
\frac{2\sqrt{2\pi}M_{Z'}}{\sqrt{\left(C^{Z'}_{L,e}\right)^2+\left(C^{Z'}_{R,e}\right)^2}}\gtrsim17.8\ \ \text{TeV},
\end{eqnarray}
for $(1+t^2)X_H>1$ which corresponds to the case $\eta_{LR},\eta_{RL}<0$. Then, we obtain a lower bound on the mass of the new gauge boson $Z'$ as
\begin{eqnarray}
M_{Z'}\gtrsim1.96(1.42)\times\frac{\left\{1+0.8(1+t^2)\left[\right(1+t^2)X_H-2]X_H\right\}^{1/2}}{t} \ \ \text{TeV},
\end{eqnarray}
for $(1+t^2)X_H<1$ $[(1+t^2)X_H>1]$.
\subsection{LHC constraint}

\begin{figure}[t]
 \centering
\begin{tabular}{cc}
\includegraphics[width=0.6 \textwidth]{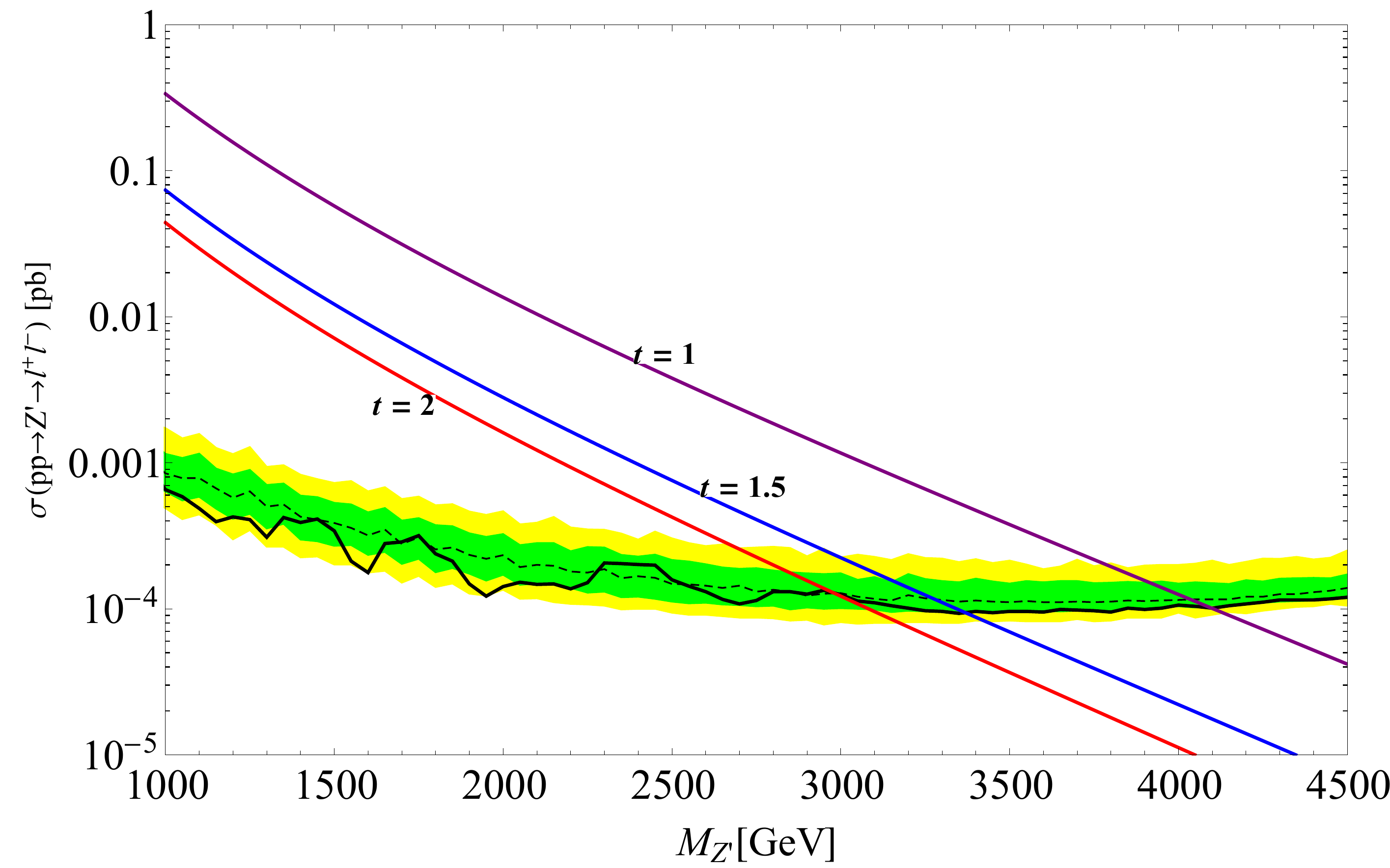}
\end{tabular}
  \caption{The cross-section for the process $pp\rightarrow Z'\rightarrow l^+l^-$ as a function of the $Z'$ gauge boson mass, at $X_H=1/6$. The solid and dashed black curves are the observed and expected limits, respectively, whereas the green and yellow bands correspond to $1\sigma$ and $2\sigma$ for the expected limit \cite{ATLAS2017}.}\label{PPtoXtoll}
\end{figure}

At LHC, the new gauge boson $Z'$ can be resonantly produced from the $q\bar{q}$ fusion and subsequently decay to the pairs of the SM fermions. The most promising channel to search for the new gauge boson $Z'$ at LHC is through Drell-Yan process, namely $pp\rightarrow Z'\rightarrow l^+l^-$ ($l=e,\mu$). The cross-section for this process at a fixed collider center-of-mass energy $\sqrt{s}$ is given, with no cut on the lepton pair rapidity, by
\begin{eqnarray}
\sigma&=&\sum_{q}\int^s_0d\hat{s}L_{q\bar{q}}(\hat{s})\hat{\sigma}(q\bar{q}\rightarrow Z'\rightarrow l^+l^-),\label{totcrrs}
\end{eqnarray}
where $\sqrt{\hat{s}}$ is the invariant mass of the dilepton system, $\hat{\sigma}(q\bar{q}\rightarrow Z'\rightarrow l^+l^-)$ is the cross-section at the partonic level given by 
\begin{eqnarray}
\hat{\sigma}(q\bar{q}\rightarrow Z'\rightarrow l^+l^-)=\frac{\hat{s}}{36\pi}\frac{\left[\left(C^{Z'}_{V,q}\right)^2+\left(C^{Z'}_{A,q}\right)^2\right]\left[\left(C^{Z'}_{V,l}\right)^2+\left(C^{Z'}_{A,l}\right)^2\right]}{\left(\hat{s}^2-M^2_{Z'}\right)^2+M^2_{Z'}\Gamma^2_{Z'}},
\end{eqnarray}
and $L_{q\bar{q}}$ is the parton luminosities defined by
\begin{eqnarray}
L_{q\bar{q}}(\hat{s})=\int^1_{\frac{\hat{s}}{s}}\frac{dx}{xs}\left[f_q(x,\hat{s})f_{\bar{q}}\left(\frac{\hat{s}}{xs},\hat{s}\right)+f_q\left(\frac{\hat{s}}{xs},\hat{s}\right)f_{\bar{q}}(x,\hat{s})\right],
\end{eqnarray}
with $f_{q(\bar{q})}(x,\hat{s})$ to be the parton distribution function of the quark $q$ (antiquark $\bar{q}$) evaluated at the scale $\hat{s}$ \cite{Stirling2009}. From the cross-section for the process $pp\rightarrow Z'\rightarrow l^+l^-$ predicted in this model and the upper limits on $\sigma\times BR$ corresponding to a new neutral gauge boson at the $95\%$ confidence level obtained by the ATLAS \cite{ATLAS2017}, we can find the constraint on the $Z'$ gauge boson mass and the gauge coupling ratio $t$ for $X_H$ kept fixed. In figure \ref{PPtoXtoll}, we show the current LHC limits and the cross-section for the process $pp\rightarrow Z'\rightarrow l^+l^-$ for various values of $t$ at $X_H=1/6$.
From this figure, one can find that the lower bound on the $Z'$ gauge boson mass is about $2.95$ TeV, $3.4$ TeV, and $4.1$ TeV corresponding to $t=2$, $t=1.5$, and $t=1$, respectively. 

\begin{figure}[t]
 \centering
\begin{tabular}{cc}
\includegraphics[width=0.6 \textwidth]{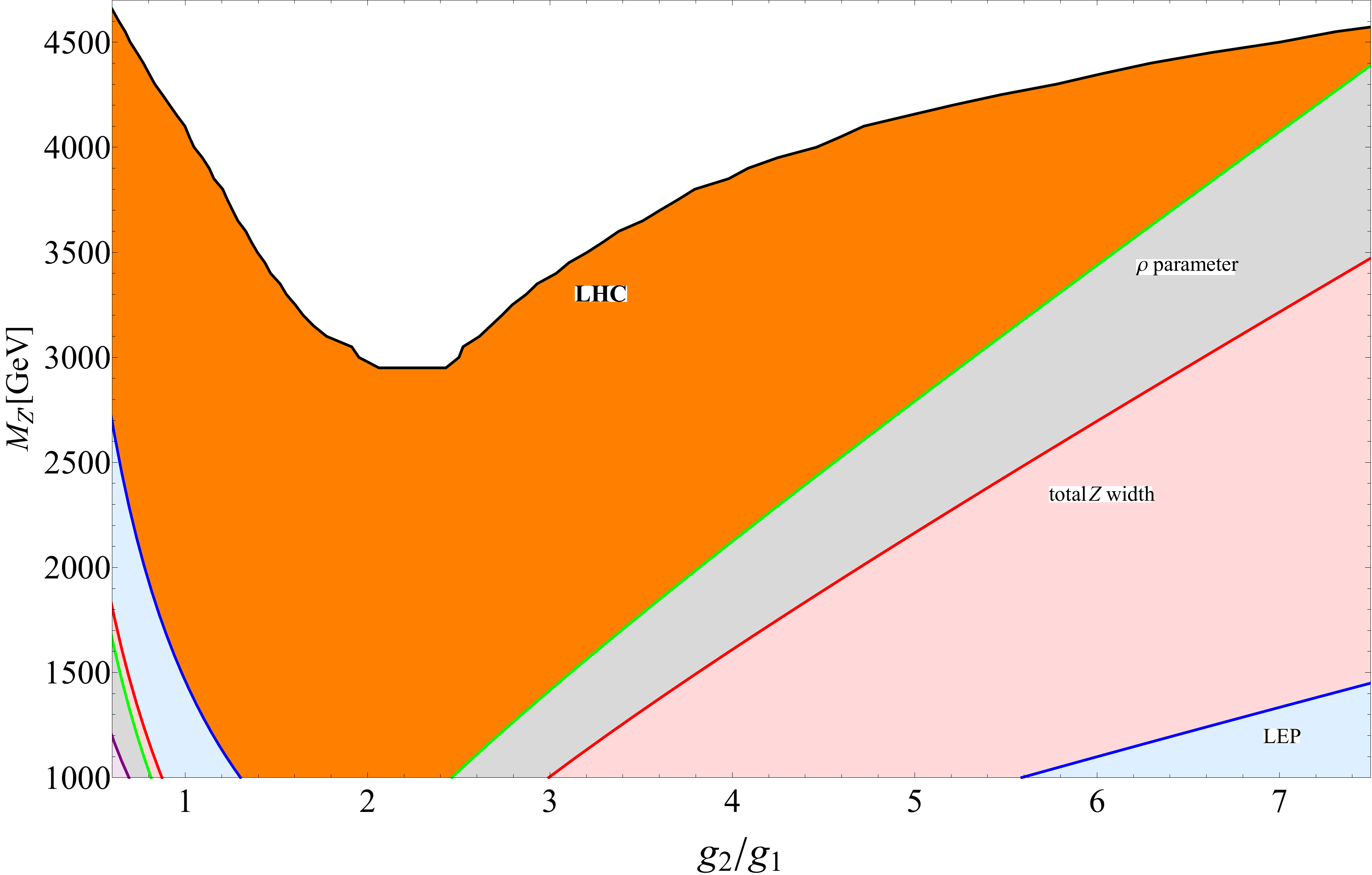}
\end{tabular}
  \caption{The allowed parameter region in the $M_{Z'}-g_2/g_1$ plane is the white region, which is obtained from the combination of the total $Z$ width, $\rho$ paramter, APV of Cesium, LEP, and current LHC constraints. Here, we use $X_H=1/6$ and the region of $g_2/g_1$ satisfies the perturbativity constraint (\ref{perconst}).}\label{MZp-t-plane}
\end{figure}

In figure \ref{MZp-t-plane}, we combine the current LHC limits and the bounds obtained above to find the allowed parameter region in the $M_{Z'}-g_2/g_1$ plane. The regions below the black, green, red, blue, and purple curves are excluded by the current LHC, $\rho$ parameter, total $Z$ width, LEP, and Cesium nuclear weak charge bounds, respectively. The white region refers to the allowed parameter space corresponding to the current experimental data.
From this figure, one can see that the direct search of new neutral gauge boson at LHC imposes the most stringent bound on the relation between $M_{Z'}$ and $g_2/g_1$. With $X_H=1/6$ and the region of $g_2/g_1$ satisfying the perturbativity constraint (\ref{perconst}), the mass of the new gauge boson $Z'$ must be constrained as, $M_Z'\gtrsim 2.9$ TeV.

\section{\label{ILC}Forward-Backward Asymmetry at ILC}
We investigate the potential of probing for the signal of the new gauge boson $Z'$ in our model by studying the deviation of forward-backward (FB) asymmetry with respect to the process $e^+e^-\rightarrow\bar{f}f$ at the International Linear Collider (ILC). The most sensitive mode corresponding to this process at ILC is
\begin{eqnarray}
e^-(k_1,\sigma_1)+e^+(k_2,\sigma_2)\rightarrow\mu^-(k_3,\sigma_3)+\mu^+(k_4,\sigma_4),
\end{eqnarray}
where $\sigma_i=\pm1$ and $k_i$ are the helicities and the $4$-momentum of the charged leptons, respectively. Since in this work, we focus on the final state mode $\mu^+\mu^-$. The helicity amplitudes for $e^+e^-\rightarrow\mu^+\mu^-$ coming from the contributions of the $\gamma$, $Z$, and $Z'$ gauge bosons are written as
\begin{eqnarray}
\mathcal{M}(+-+-)&=&-4\pi\alpha(1+\cos\theta)s\left[\frac{1}{s}+\frac{t^2_W}{s_Z}+\frac{(C^{Z'}_{R,e})^2}{4\pi\alpha M^2_{Z'}}\right],\nonumber\\
\mathcal{M}(-+-+)&=&-4\pi\alpha(1+\cos\theta)s\left[\frac{1}{s}+\frac{(-\cot2\theta_W)^2}{s_Z}+\frac{(C^{Z'}_{L,e})^2}{4\pi\alpha M^2_{Z'}}\right],\nonumber\\
\mathcal{M}(+--+)&=&\mathcal{M}(-++-)=4\pi\alpha(1-\cos\theta)s\left[\frac{1}{s}-\frac{t_W\cot2\theta_W}{s_Z}+\frac{C^{Z'}_{R,e}C^{Z'}_{L,e}}{4\pi\alpha M^2_{Z'}}\right],\nonumber\\
\mathcal{M}(++++)&=&\mathcal{M}(----)=0,
\end{eqnarray}
where $\theta$ is the scattering polar angle, $s=(k_1+k_2)^2=(k_3+k_4)^2$, $s_Z=s-M^2_Z+iM_Z\Gamma_Z$, 
and $\alpha=e^2/4\pi$ is the fine-structure constant.

In general, the partially-polarized differential cross-section is defined as follows \cite{Verzegnassi1992}
\begin{eqnarray}
\frac{d\sigma(P_{e^-},P_{e^+})}{d\cos\theta}=\sum_{\sigma_1,\sigma_2}\frac{1+\sigma_1 P_{e^-}}{2}\frac{1+\sigma_2 P_{e^+}}{2}\frac{d\sigma_{\sigma_1\sigma_2}}{d\cos\theta},
\end{eqnarray}
where $P_{e^-}$ and $P_{e^+}$ are the degrees of polarization for the electron and positron beams, respectively, and $d\sigma_{\sigma_1\sigma_2}/d\cos\theta$ is the differential cross-section for purely-polarized initial state with the helicity of the final states summed up, given by
\begin{eqnarray}
\frac{d\sigma_{\sigma_1\sigma_2}}{d\cos\theta}=\frac{1}{32\pi s}\sum_{\sigma_3,\sigma_4}|\mathcal{M}|^2.
\end{eqnarray}
By following the realistic values at ILC \cite{ILC}, one define the polarized differential cross-sections as
\begin{eqnarray}
\frac{d\sigma_R}{d\cos\theta}\equiv\frac{d\sigma(0.8,-0.3)}{d\cos\theta}, \ \ \ \ \frac{d\sigma_L}{d\cos\theta}\equiv\frac{d\sigma(-0.8,0.3)}{d\cos\theta}.\label{ILCpol}
\end{eqnarray}
Then, the forward-backward asymmetry is determined by the following quantity
\begin{eqnarray}
A^i_{FB}=\frac{N^i_F-N^i_B}{N^i_F+N^i_B},
\end{eqnarray}
where the index $i$ refers to $L$ or $R$, and the number of the forward (backward) events is defined by
\begin{eqnarray}
N^{i}_{F(B)}=\epsilon\mathcal{L}\int^{c_{\text{max}}(0)}_{0(-c_{\text{max}})}d\cos\theta\frac{d\sigma_i}{d\cos\theta},
\end{eqnarray}
with $\epsilon$ to be the efficiency of observing the events which is equal to one for electron and muon final states, $c_{\text{max}}=0.95$ with respect to the muon final state \cite{Tran2016}. 

The sensitivity to the contribution of the new gauge boson $Z'$ in FB asymmetry of the process $e^+e^-\rightarrow\mu^+\mu^-$ is determined by the following quantity 
\begin{eqnarray}
\Delta A^{L(R)}_{FB}=\left|A^{L(R)}_{FB}\Big|_{\text{SM}+Z'}-A^{L(R)}_{FB}\Big|_{\text{SM}}\right|,
\end{eqnarray}
where $A^{L(R)}_{FB}\Big|_{\text{SM}}$ and $A^{L(R)}_{FB}\Big|_{\text{SM}+Z'}$ represent FB asymmetry for the cases coming from only the SM boson contribution and from the SM plus $Z'$ boson contributions. This quantity should be compared with the statistical error of FB asymmetry associated with only the SM boson contribution given by \cite{Verzegnassi1992,Okada2018}
\begin{eqnarray}
\delta A^{L(R)}_{FB}=\sqrt{\frac{1-\left(A^{L(R)}_{FB}\Big|_{\text{SM}}\right)^2}{N^{L(R)}_F\Big|_{\text{SM}}+N^{L(R)}_B\Big|_{\text{SM}}}}. 
\end{eqnarray}
\begin{figure}[t]
 \centering
\begin{tabular}{cc}
\includegraphics[width=0.45 \textwidth]{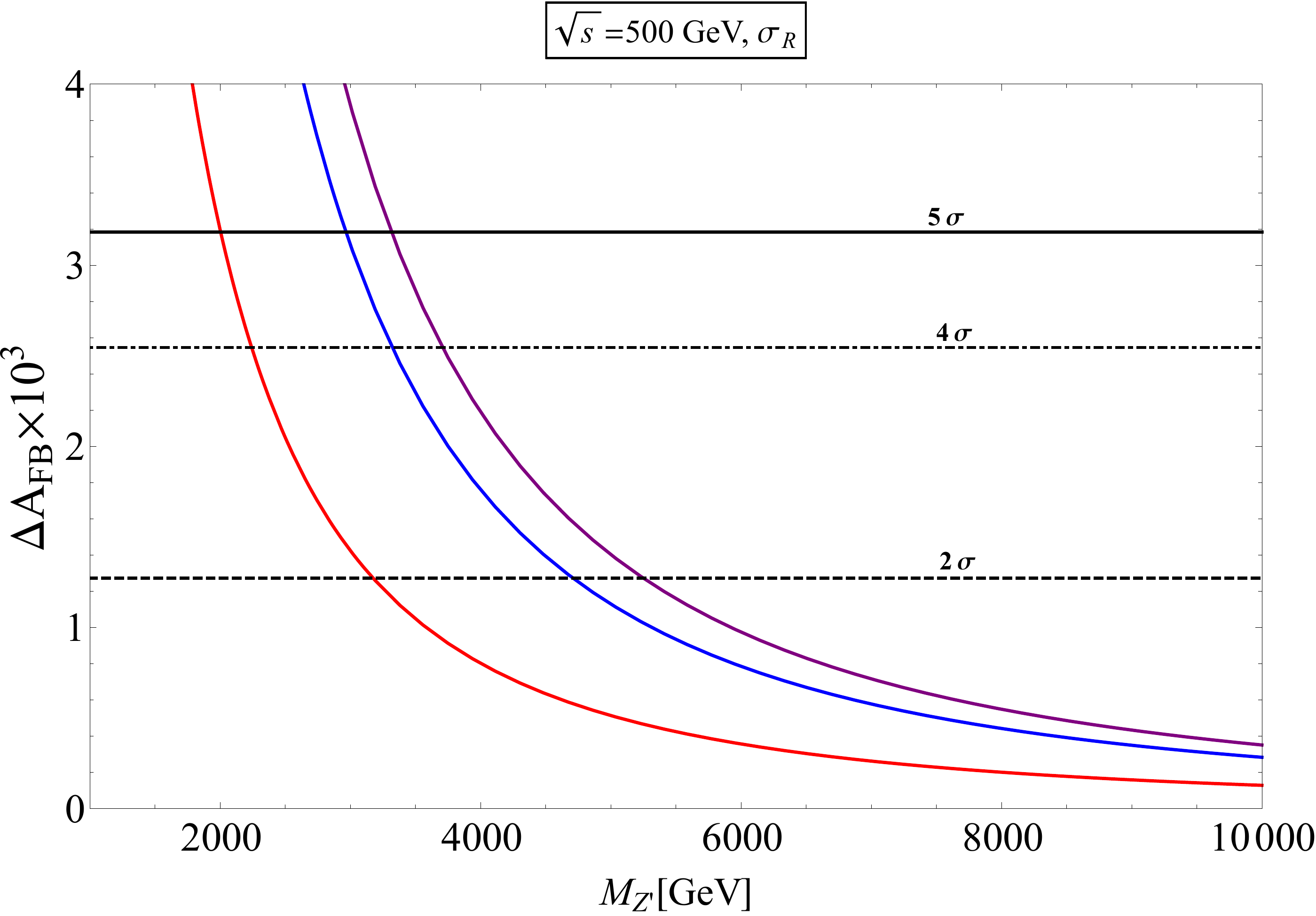}
\hspace*{0.05\textwidth}
\includegraphics[width=0.45 \textwidth]{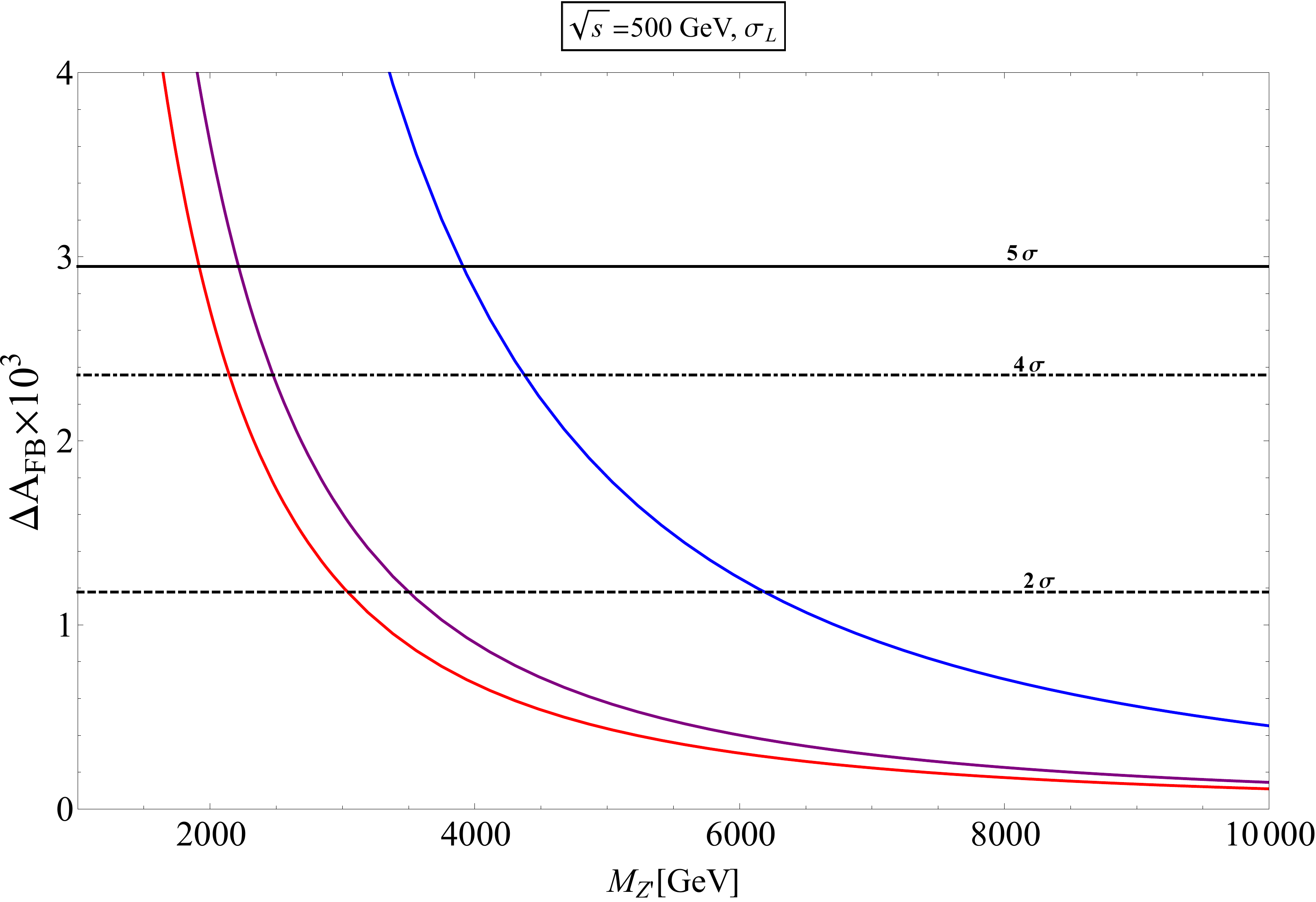}\\
\includegraphics[width=0.45 \textwidth]{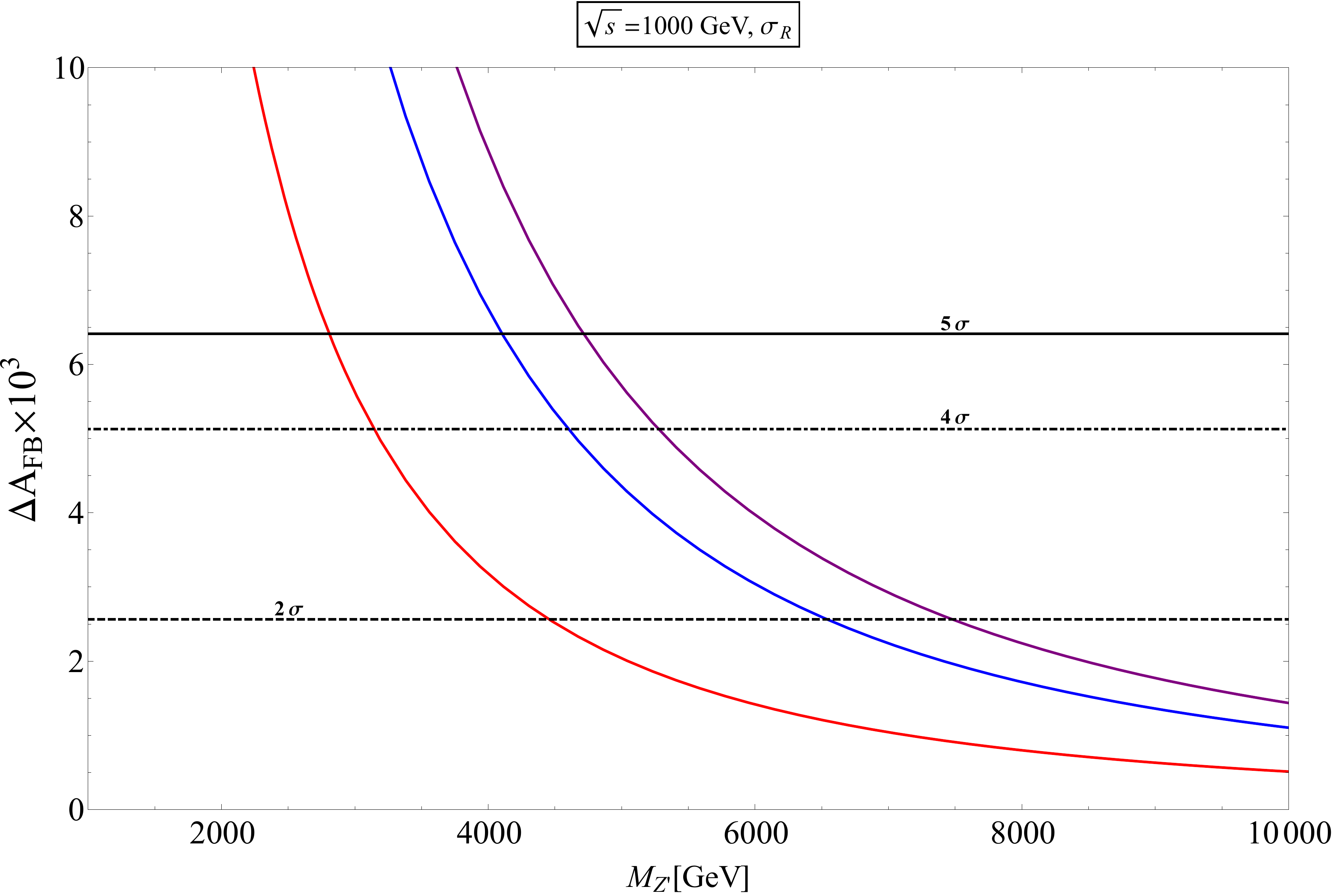}
\hspace*{0.05\textwidth}
\includegraphics[width=0.45 \textwidth]{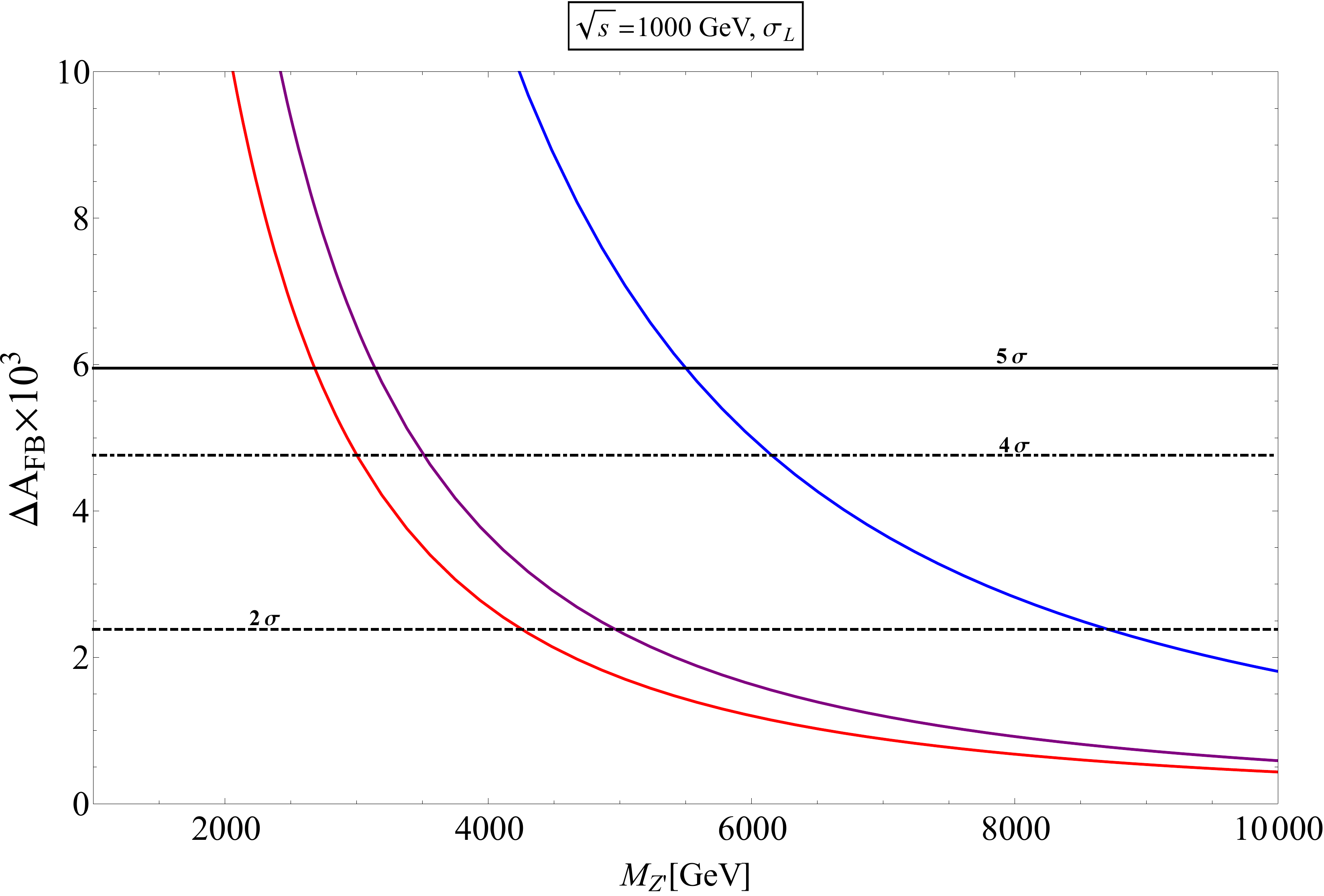}
\end{tabular}
  \caption{The quantity $\Delta A^{L(R)}_{FB}$, describing the contribution of the new gauge boson $Z'$ to FB assymetry for the process $e^+e^-\rightarrow\mu^+\mu^-$ at ILC, as a function of $M_{Z'}$. The blue, red, and purple curves correspond to $t=1$, $1.5$, and $4$, respectively. The horizontal lines correspond to the confidence levels $2\sigma$, $4\sigma$, and $5\sigma$, respectively. Here, we set $X_H=1/6$ and $\mathcal{L}=4000$ $\text{fb}^{-1}$.}\label{AFB-ILC}
\end{figure}
We estimate the sensitivity to the contribution of the new gauge boson $Z'$ in FB asymmetry of the process $e^+e^-\rightarrow\mu^+\mu^-$ by requiring $\Delta A^{L(R)}_{FB}>2\sigma$ which the $Z'$ signal could manifest itself over the SM background. 

\begin{figure}[t]
 \centering
\begin{tabular}{cc}
\includegraphics[width=0.45 \textwidth]{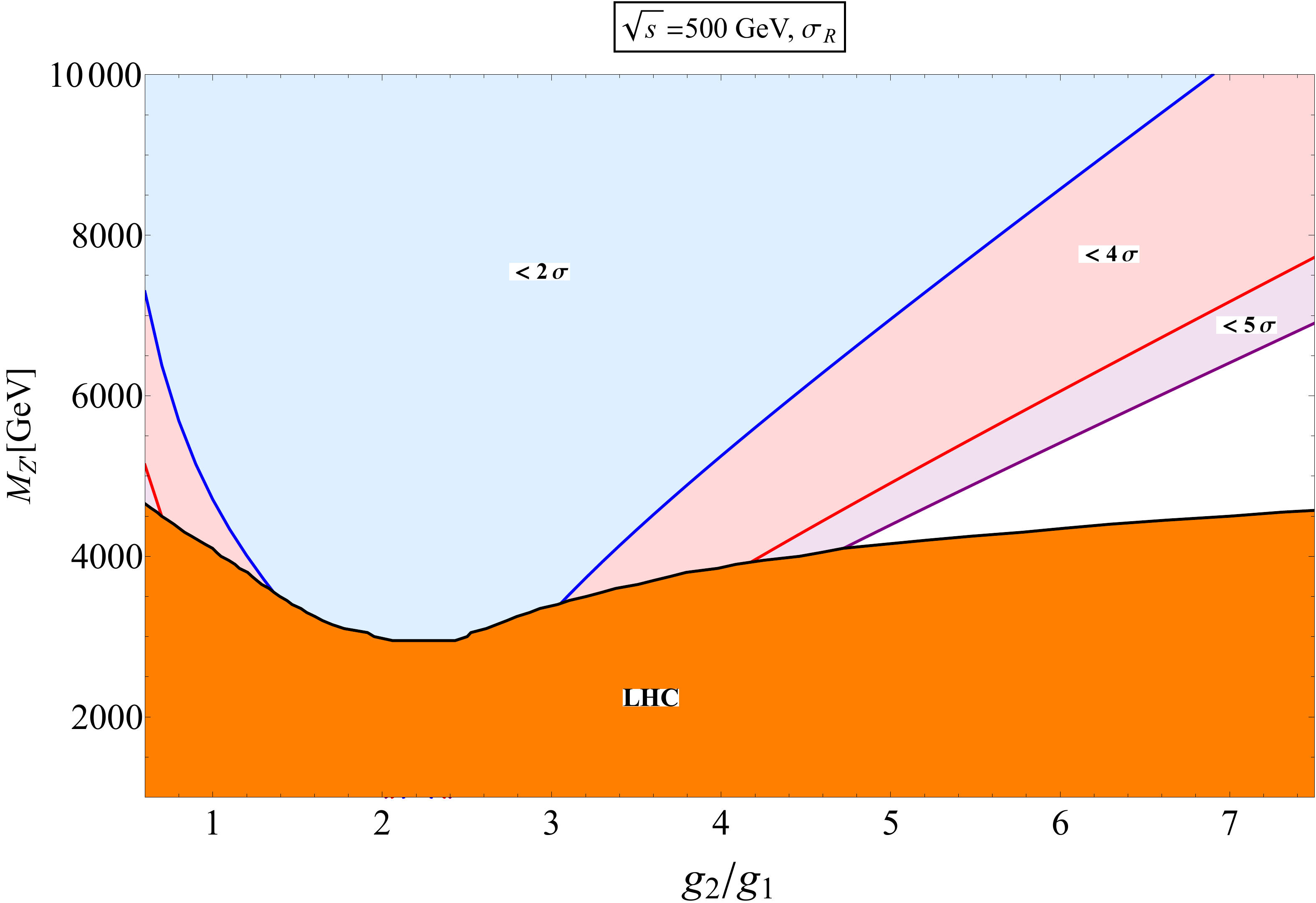}
\hspace*{0.05\textwidth}
\includegraphics[width=0.45 \textwidth]{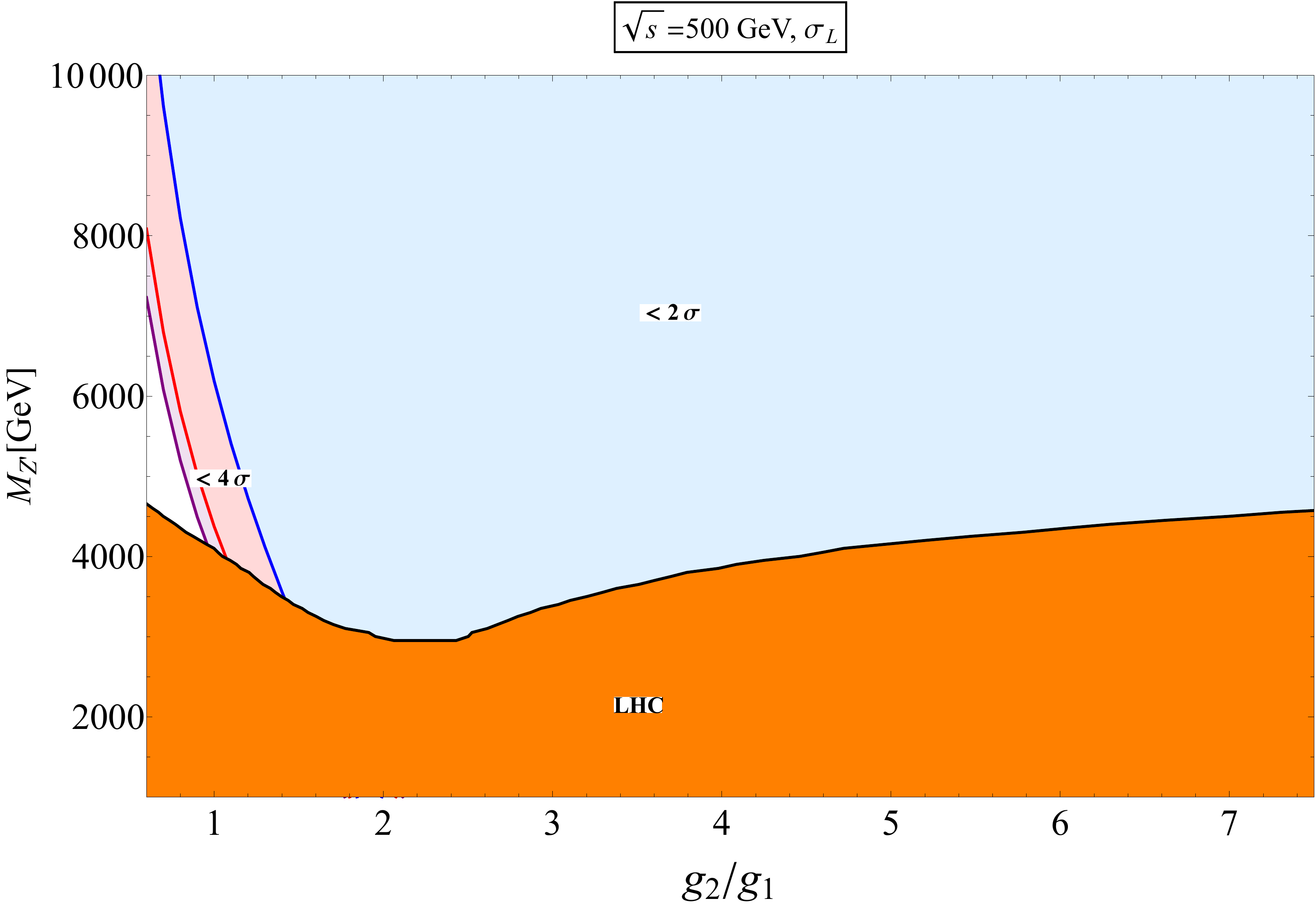}\\
\includegraphics[width=0.45 \textwidth]{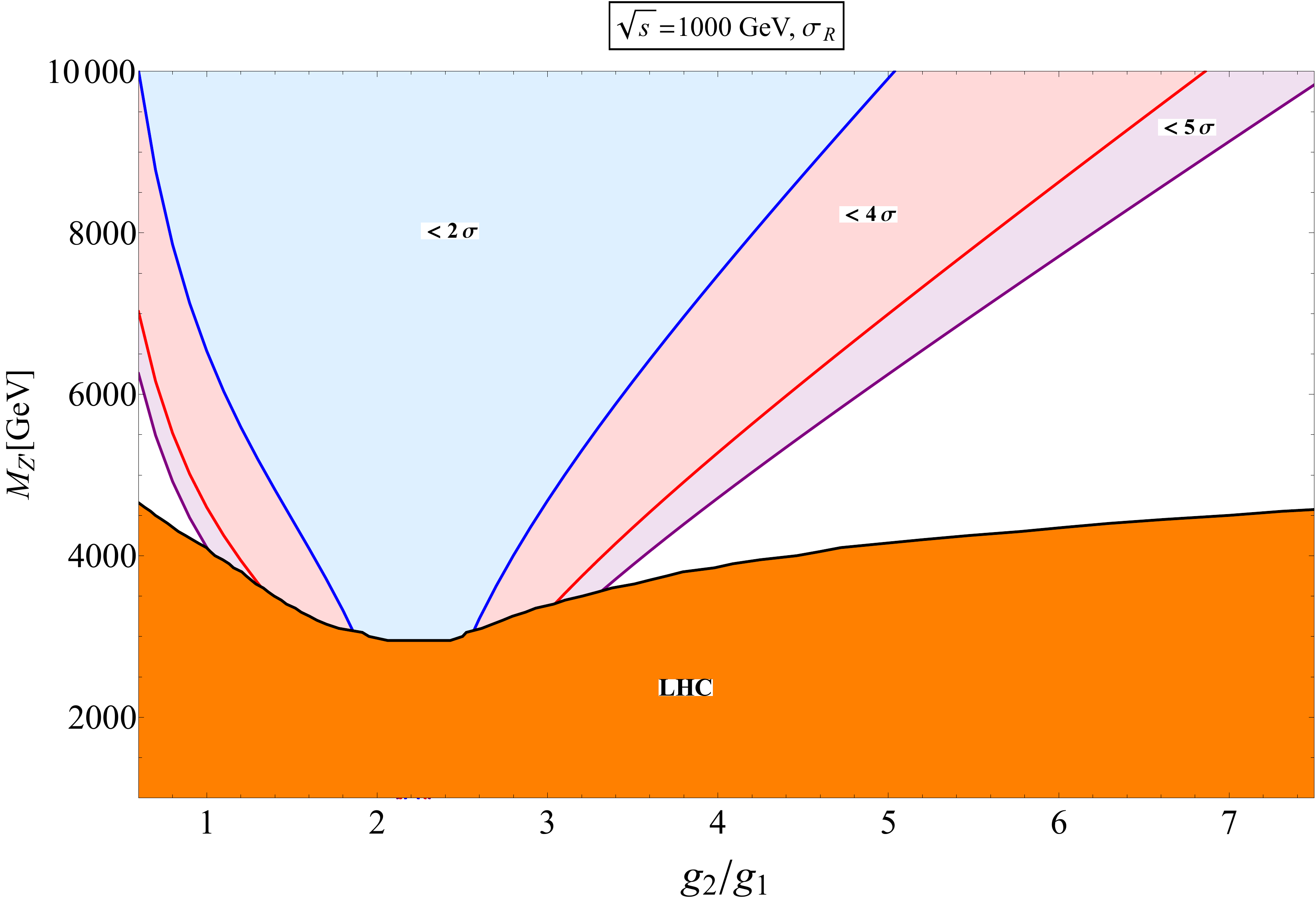}
\hspace*{0.05\textwidth}
\includegraphics[width=0.45 \textwidth]{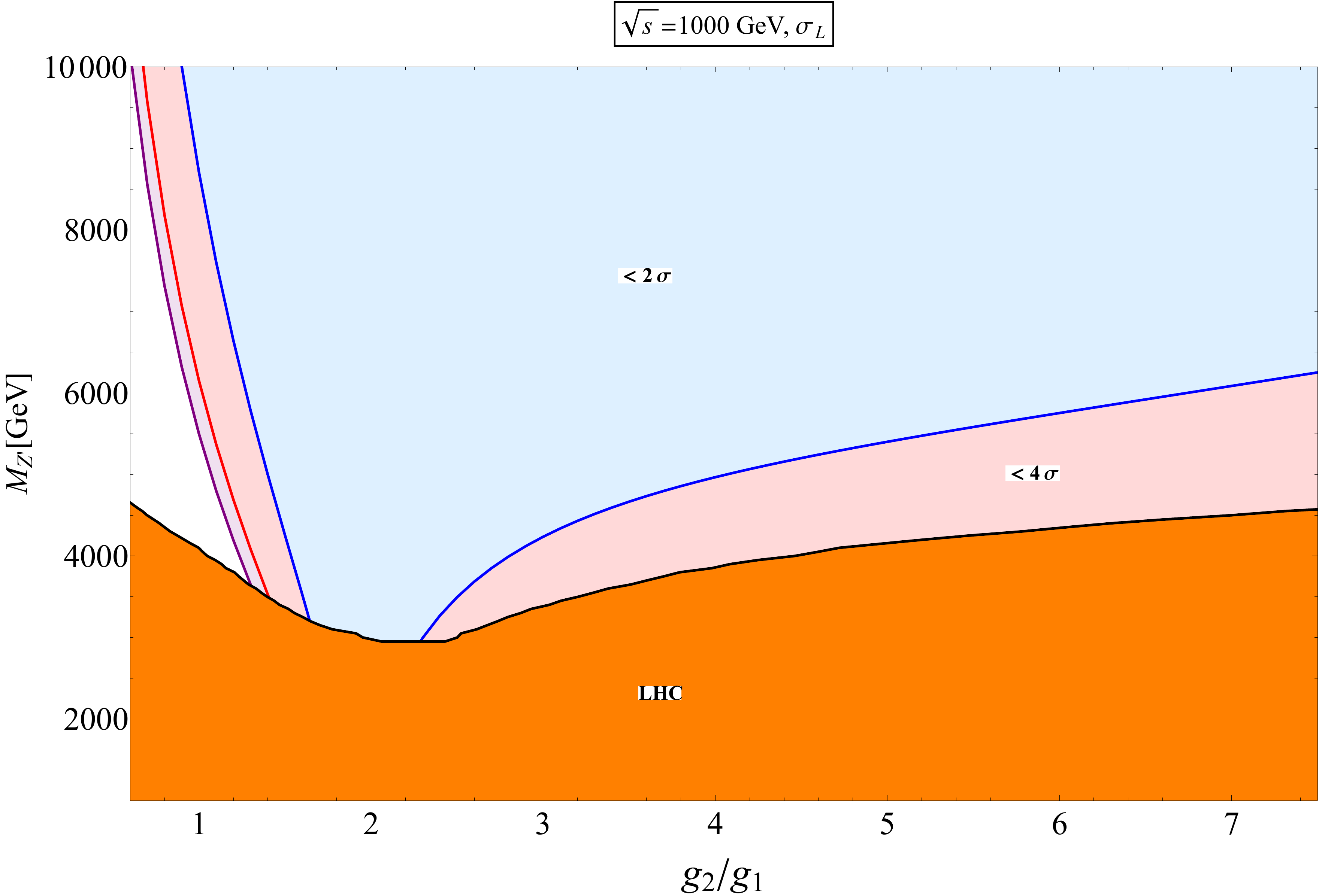}\\
\end{tabular}
  \caption{The parameter regions in the $M_{Z'}-g_2/g_1$ plane can give the $>2\sigma$ sensitivity, the $\geq4\sigma$ sensitivity, and discovery reach at $\geq5\sigma$ statistical significance, at $X_H=1/6$ and $\mathcal{L}=4000$ $\text{fb}^{-1}$.}\label{AFB-APS}
\end{figure}

In figure \ref{AFB-ILC}, we show the quantity $\Delta A^{L(R)}_{FB}$ as a function of $M_{Z'}$, for the polarized cross-sections $\sigma_{R,L}$ and various values of $t$ and $\sqrt{s}$, at integrated luminosity $\mathcal{L}=4000$ $\text{fb}^{-1}$. From this figure and the allowed parameter region obtained in the previous section, we find the regions which can give the $>2\sigma$ sensitivity, the $\geq4\sigma$ sensitivity, and discovery reach at $\geq5\sigma$ statistical significance, given in table \ref{AFB-tab}.
\begin{table}[!htp]
\centering
\begin{tabular}{|c|c|c|c|c|c|c|}
  \hline
  \multicolumn{7}{|c|}{$\sqrt{s}=500$ GeV}\\
  \hline
  & \multicolumn{3}{|c|}{$\sigma_R$} & \multicolumn{3}{|c|}{$\sigma_L$} \\
  \hline
  & $t=1$ & $t=1.5$ & $t=4$ & $t=1$ & $t=1.5$ & $t=4$ \\
  \hline
  $>2\sigma$ & $4.1\lesssim\frac{M_{Z'}}{\text{TeV}}\lesssim4.7$ & $-$ & $3.9\lesssim\frac{M_{Z'}}{\text{TeV}}\lesssim5.3$ &    
  $4.1\lesssim\frac{M_{Z'}}{\text{TeV}}\lesssim6.2$ & $-$ & $-$\\
  \hline 
  $\geq4\sigma$ & $-$ & $-$ & $-$ &    
  $4.1\lesssim\frac{M_{Z'}}{\text{TeV}}\lesssim4.4$ & $-$ & $-$\\
 \hline
 $\geq5\sigma$ & $-$ & $-$ & $-$ &    
  $-$ & $-$ & $-$\\
 \hline
 \multicolumn{7}{|c|}{$\sqrt{s}=1000$ GeV}\\
  \hline
  & \multicolumn{3}{|c|}{$\sigma_R$} & \multicolumn{3}{|c|}{$\sigma_L$} \\
  \hline
  & $t=1$ & $t=1.5$ & $t=4$ & $t=1$ & $t=1.5$ & $t=4$ \\
  \hline
  $>2\sigma$ & $4.1\lesssim\frac{M_{Z'}}{\text{TeV}}\lesssim6.5$ & $3.3\lesssim\frac{M_{Z'}}{\text{TeV}}\lesssim4.5$ & $3.9\lesssim\frac{M_{Z'}}{\text{TeV}}\lesssim7.5$ &    
  $4.1\lesssim\frac{M_{Z'}}{\text{TeV}}\lesssim8.7$ & $3.3\lesssim\frac{M_{Z'}}{\text{TeV}}\lesssim4.3$ & $3.9\lesssim\frac{M_{Z'}}{\text{TeV}}\lesssim5$ \\
  \hline 
  $\geq4\sigma$ & $4.1\lesssim\frac{M_{Z'}}{\text{TeV}}\lesssim4.6$ & $-$ & $3.9\lesssim\frac{M_{Z'}}{\text{TeV}}\lesssim5.3$ &    
  $4.1\lesssim\frac{M_{Z'}}{\text{TeV}}\lesssim6.2$ & $-$ & $-$\\
 \hline
 $\geq5\sigma$ & $-$ & $-$ & $3.9\lesssim\frac{M_{Z'}}{\text{TeV}}\lesssim4.7$ & $4.1\lesssim\frac{M_{Z'}}{\text{TeV}}\lesssim5.5$ & $-$ & $-$\\
 \hline
\end{tabular}
\caption{The regions can give the $>2\sigma$ sensitivity, the $\geq4\sigma$ sensitivity, and discovery reach at $\geq5\sigma$ statistical significance for various values of $t$ and $\sqrt{s}$, at $X_H=1/6$ and $\mathcal{L}=4000$ $\text{fb}^{-1}$.}\label{AFB-tab}
\end{table}
In more general way, we present the regions which can give these confidence levels in figure \ref{AFB-APS}.
The region which can give the $>2\sigma$ sensitivity is above the black curve but below the blue one. Whereas, the regions which can give the $\geq4\sigma$ sensitivity and discovery reach at $\geq5\sigma$ statistical significance are above the black curve but at/below the red and purple curves, respectively. From this figure, we see that at the available colliding energy and the sufficient integrated luminosity, FB assymetry is quite sensitive for the polarized cross-section $\sigma_R$ at the sufficient large $t$ region. Whereas, with both of the sufficient colliding energy and integrated luminosity, FB assymetry for $\sigma_R$ is sensitive at both of the sufficient small and large $t$ region. For the polarized cross-section $\sigma_L$, the corresponding FB assymetry is relatively less sensitive at the sufficient large $t$ region.

\section{\label{conclu} Conclusion}
So far, the $U(1)_R$ extensions of the Standard Model (SM), where only the right-handed fermions are charged under $U(1)_R$, are based on the gauge symmetry $SU(3)_C\times SU(2)_L\times U(1)_Y\times U(1)_R$. In this paper, we have proposed another $U(1)_R$ extension of the SM in the framework of the gauge symmetry $SU(3)_C\times SU(2)_L\times U(1)_{Y'}\times U(1)_R$ where the $U(1)_Y$ weak hypercharge of the SM is identified as a combination of the $U(1)_{Y'}$ and $U(1)_R$ charges. The gauge symmetry is spontaneously broken through two stages as, $SU(2)_L\times U(1)_{Y'}\times U(1)_R\rightarrow SU(2)_L\times U(1)_Y\rightarrow U(1)_{\text{em}}$. We have determined the gauge charges of the fields from the conditions of the anomaly cancellation and the gauge invariance of the Yukawa couplings. We have also indicated that the light neutrino masses can be explained through the type-I seesaw mechanism where the Majorana masses of the right-handed neutrinos are related to the $U(1)_{Y'}\times U(1)_R$ symmetry breaking scale. 

We have studied the two-body decays of the new neutral gauge boson at the tree level. We have found that the dominant decay channels, which are either the dilepton or dijet modes, depend on the values of the free parameters of the model. In addition, we have obtained the constraints on the free parameters of the model based on precision measurement of the total $Z$ width, perturbativity condition, $\rho$ parameter, atomic parity violation of Cesium, the $e^+e^-\rightarrow l^+l^-$ process at LEP, and the current LHC limits on the production of new gauge boson. Finally, in order to test the model, we have investigated the forward-backward asymmetry for the process $e^+e^-\rightarrow\mu^+\mu^-$ which is the most sensitive mode at ILC. We have found that, for the colliding energy and integrated luminosity which are expected in the upgraded ILC, there are the significant parameter regions which can give the $>2\sigma$, $\geq4\sigma$ sensitivities and discovery reach at $\geq5\sigma$ statistical significance for the signal of the new neutral gauge boson.

\end{document}